  \documentclass{emulateapj}

\usepackage{graphicx,natbib}
\usepackage{hyperref}

\newcommand{\ms}{M$_{\odot}$}

\shorttitle{Apsidal motion and LC solution for 13 LMC eccentric eclipsing binaries}
\shortauthors{Zasche et al.}

\begin{document}

\title{Apsidal motion and a light curve solution \\ for thirteen LMC eccentric eclipsing binaries\thanks{Based
  on data collected with the Danish 1.54-m telescope at the ESO La Silla Observatory.}}

\author{P. Zasche\altaffilmark{1}, M. Wolf\altaffilmark{1}, J. Vra\v{s}til\altaffilmark{1}, L. Pilar\v{c}\'{\i}k\altaffilmark{1}}
 \affil{
 \altaffilmark{1} Astronomical Institute, Charles University in Prague, Faculty of
Mathematics and Physics,\\ CZ-180 00 Praha 8, V Hole\v{s}ovi\v{c}k\'ach 2, Czech Republic }


\begin{abstract}
\noindent New CCD observations for thirteen eccentric eclipsing binaries from the Large Magellanic
Cloud were carried out using the Danish 1.54-meter telescope located at the La Silla Observatory in
Chile. These systems were observed for their times of minima and 56 new minima were obtained. These
are needed for accurate determination of the apsidal motion. Besides that, in total 436 times of
minima were derived from the photometric databases OGLE and MACHO.
 {The $O-C$\ diagrams of minima timings for these B-type binaries were analysed and the parameters
of the apsidal motion were computed. The light curves of these systems were fitted using the
program {\sc PHOEBE}, giving the light curve parameters.}
 {We derived for the first time the relatively short periods of the apsidal motion ranging from
21 to 107 years. The system OGLE-LMC-ECL-07902 was also analysed using the spectra and radial
velocities, resulting in masses 6.8, and 4.4 $M_\odot$ for the eclipsing components. For one system
(OGLE-LMC-ECL-20112), the third-body hypothesis was also used for describing the residuals after
subtraction of the apsidal motion, resulting in period of about 22~years. For several systems an
additional third light was also detected, which makes these systems suspicious of triplicity.}
\end{abstract}


\section{Introduction}

In some aspects the focus of astronomers to eclipsing binaries moved from the galactic to
extragalactic targets. This is mainly due to the large and long-lasting photometric monitoring
projects. Such surveys like MACHO or OGLE have discovered many thousands of new eclipsing binaries
in both Large and Small Magellanic Clouds, hence, we know only about twice more eclipsing binaries
in our own Milky Way than in other galaxies (see Pawlak et al. 2013, or Graczyk et al. 2011).    

The role of eclipsing binaries to our current astrophysical knowledge is undisputable. For example
the eccentric eclipsing binaries (hereafter EEBs) with an apsidal motion can provide us with an
important observational test of theoretical models of stellar structure and evolution. A long-term
collection of the times of EEBs minima observed for several years during its apsidal motion cycle
and their analysis can provide us with both the orbital eccentricity and the period of rotation of
the apsidal line with high accuracy (Gim\'enez 1994). Many different sets of stellar evolution
models have been published in recent years, such as Maeder (1999), Claret (2005), the MESA code
(Paxton et al. 2011), the Y$^2$ models (Demarque et al. 2004), or others. However, to distinguish
between them and to test, which one is more suitable, it is still rather difficult (see e.g.
Martins \& Palacios 2013). The internal structure constants, as derived from the apsidal motion
analysis, could serve as one independent criterion which can be used. However, to discriminate
between the models one would need an accuracy of internal structure constants of about 1\,\%, which
can be achieved only with very precise photometric and spectroscopic data.

However, the chemical composition of the Magellanic Clouds differs a bit from that of our solar
neighborhood (see e.g. Westerlund 1997, or Davies et al. 2015), and the study of the massive and
metal-deficient stars in the Magellanic Clouds checks our evolutionary models for these abundances.
All of the eclipsing binaries analysed in the present study have properties that make them
important astrophysical laboratories for studying the stellar structure and evolution of massive
stars (Ribas 2004).

In the following sections we analyse the photometric data obtained during the automatic survey as
well as our own observations and derive the rates of apsidal motion for thirteen detached eclipsing
systems with eccentric orbits located in the Large Magellanic Cloud. All these systems are
early-type objects, exhibit an apsidal motion, and was only poorly studied until now. Similar
studies of LMC EEBs have been presented by Michalska \& Pigulski (2005), by Michalska (2007), by
Zasche \& Wolf (2013), and recently also by Hong et al. (2014). A set of such binaries with known
apsidal motion is still rather limited outside of our Galaxy, hence a new contribution to the topic
is still very important. It can also serve as a testing benchmark or a starting point for some
future more detailed investigation of such systems.

\section{Observations of minimum light}

The analysis of mid-eclipse times measurements became a popular method nowadays, especially thanks
to the superb precision and a time coverage of the Kepler targets, see e.g. Borkovits et al.
(2015). Moreover, such a photometric monitoring of faint EEBs in external galaxies became almost
routine nowadays with quite modest telescopes of 1 - 2m class, which are equipped with a modern CCD
camera. However, a large amount of observing time is needed, which is usually unavailable at larger
telescopes. This especially apply for spectroscopy and other more challenging techniques, for the
photometry the situation is much more promising.

Therefore, during the last three observational seasons, we have accumulated many photometric
observations and derived 56 precise times of minimum light for selected eccentric systems. The
systems for our presented analysis were chosen following an easy criterion: all of them were
already observed during the previous seasons in the fields of the already known apsidal motion
stars in the LMC. New CCD photometry was obtained at the La Silla Observatory in Chile, where the
1.54-m Danish telescope (hereafter DK154) with the CCD camera and \textit{R} filter was used
(remotely operated from the Czech Republic).

A standard procedure for the reduction was used, applying the bias frames and then the flat fields
to the CCD frames. The comparison star was chosen to be close to the variable one and with similar
spectral type. A custom-made aperture-photometry reduction software {\sc Aphot} developed by
M.~Velen and P.~Pravec, was routinely used for reducing the data. No correction for differential
extinction was applied because of the proximity of the comparison stars to the variable and the
resulting negligible differences in air mass and their similar spectral types.

The new times of primary and secondary minima and their respective errors were determined by the
classical Kwee-van Woerden (1956) method or by our new approach (see Zasche et al. 2014). All new
times of minima are given in the online-only appendix Table \ref{minimaTable}.

\section{Photometry and light curve modelling} \label{modelling}

The main part of our present analysis lies on the huge photometric data sets as obtained during the
{\sc MACHO} and {\sc OGLE} surveys. From these large surveys the photometric catalogues with
thousands of eclipsing binaries were used for our analysis, namely: the {\sc Macho} (Faccioli et
al. 2007), {\sc Ogle II} (Wyrzykowski et al. 2004), and {\sc Ogle III} (Graczyk et al. 2011)
databases. These data were used both for analysing the minima times as well as for the analysis of
the light curve. Our new observations carried out with the DK154 were used only to derive the times
of minimum light for the studied systems because only small parts of the light curves near the
minima were observed.

The light curve (hereafter LC) analysis for a particular system was carried out using the program
{\sc PHOEBE}, ver. 0.31a (Pr{\v s}a \& Zwitter 2005), which is based on the Wilson-Devinney
algorithm (Wilson \& Devinney 1971) and its later modifications. However, some of the parameters
have to be fixed during the fitting process. The albedo coefficients $A_i$ remained fixed at a
value 1.0, the gravity darkening coefficients $g_i = 1.0$. The limb darkening coefficients were
interpolated from the van Hamme's tables (van Hamme 1993), and the synchronicity parameters ($F_i$)
were also kept fixed at values of $F_i = 1$. The temperature of the primary component was derived
from the photometric indices. The $(B-V)_0$ and $(V-I)_0$ indices were used for the estimation of
the primary temperature (see the next paragraph), see the Table \ref{InfoSystems}. Hence, the set
of parameters derived during the LC fitting of all the systems is the following: secondary
temperature $T_2$, inclination $i$, the Kopal's modified potentials $\Omega_i$, and the
luminosities $L_i$.

The proper primary temperature was mostly found by using the photometric indices as published by
the OGLE team (see Table \ref{InfoSystems}), or by using the $UBV$ magnitudes as published by
Massey (2002), or Zaritsky et al. (2004). After then, the dereddened indices (using the method by
Johnson 1958) where used for estimation of a spectral type and roughly also its temperature (e.g.
from Pecaut \& Mamajek 2013). The similar method was also used for the $(V-I)_0$ indices, while the
$E(V-I)$ was taken as an average of the $E(V-I)$ values given by Ulaczyk et al. (2012). All of
these binaries seem to be of B1-B9 spectral type.

The problematic issue of the mass ratio (without having any spectroscopy) was solved in the
following way. Because the detached eclipsing binaries and their LC solution is only poorly
sensitive to the photometric mass ratio (see e.g. Terrell \& Wilson 2005), we used a different
approach as described e.g. in Graczyk (2003). In this approach, we assume that the stars follow a
standard mass-luminosity relation and hence the estimated mass ratio can be derived from the
following equation: $$ q = 10^{(\log L2 - \log L1)/3.664}, $$ where the luminosity values of the
two components were taken from the LC solution from {\sc PHOEBE}. Using this method iteratively, we
obtain a new photometric mass ratio after only a few steps (usually three to five). Such a value of
$q$ is physically self-consistent with the derived values of radii, luminosities, etc.

Here we would like to point out that the solution found with the {\sc PHOEBE} program is a formal
one, based on the abovementioned assumptions (the inclusion of some future knowledge of these stars
can significantly shift our solution). Moreover, also the errors as resulted from the code are
purely mathematical ones and usually are strongly underestimated in the {\sc PHOEBE} program (Pr{\v
s}a \& Zwitter 2005, or PHOEBE
manual\footnote{http://phoebe-project.org/1.0/docs/phoebe$\_$manual.pdf}).

\section{Apsidal motion analysis}

For the analysis of period changes using the times of minima, we used the approach as presented
below.

 \begin{enumerate}
   \item At first, all photometric data were analysed, resulting in a set of preliminary minima
   times. Hence, also some preliminary apsidal motion parameters were derived (with the assumption $i=90^\circ$).\\[-1mm]

   \item At second, the eccentricity ($e$), argument of periastron ($\omega$), and the apsidal motion
   rate ($\dot \omega$) that resulted from the apsidal motion analysis were used for the preliminary
   light curve analysis.\\[-1mm]

   \item Then the inclination ($i$) from the LC analysis was used for the final apsidal motion
   analysis.\\[-1mm]

   \item And finally, the resulted $e$, $\omega$, and $\dot \omega$ values from the apsidal motion
   analysis were used for the final LC analysis.\\[-1mm]
 \end{enumerate}

In general, the differences between the preliminary results and the final ones as resulted from
steps 1 and 3, and 2 and 4 are usually rather small. Moreover, this approach was a bit complicated
because the minima times were also derived using the light curve template. Hence, the LC solution
from step 2 allows us to derive the better times of minima for the step 3. The whole process runs
iteratively until the changes are negligible (usually after running these four steps twice).

All of the times of minima were analysed using the method presented by Gim\'enez \&
Garc\'{\i}a-Pelayo (1983). This is a least-squares iterative procedure, including terms in the
eccentricity up to the fifth order. There were five independent variables $(T_0, P_s, e,
\dot{\omega}, \omega_0)$ derived. The argument of periastron $\omega$ is then given by the linear
equation $\omega = \omega_0 + \dot{\omega}\ E,$ where $\dot{\omega}$ is the rate of periastron
advance, $E$ is the epoch, and the position of periastron for the zero epoch $T_0$ is denoted as
$\omega_0$. The relation between the sidereal and the anomalistic periods, $P_s$ and $P_a$, is then
given by

 \medskip
\noindent $ P_s = P_a \,(1 - \dot{\omega}/360^\circ) $

 \medskip
\noindent and the period of the apsidal motion is $ U = 360^\circ P_a/\dot{\omega} $.

For all of the minima the individual weights were derived from their respective uncertainties. All
of these data are stored in the online-only appendix (for a sample see Table \ref{minimaTable}).

\begin{table}[b]
 \centering
  \begin{minipage}{88mm}
 \fontsize{1.8mm}{2.4mm}\selectfont
 \caption{List of the minima timings used for the analysis.} \label{minimaTable}
\begin{tabular}{ccclcl}
\hline\hline\noalign{\smallskip}
 Star       &    JD Hel.- &  Error & Type   &  Filter  & Source /     \\
            &   2400000   &  [day] &        &          & Observatory  \\
\noalign{\smallskip}\hline \noalign{\smallskip}
 OGLE-LMC-ECL-07902 & 48750.43767 & 0.00180 & Prim & B+R & MACHO \\
 OGLE-LMC-ECL-07902 & 48751.26899 & 0.00262 & Sec  & B+R & MACHO \\
 OGLE-LMC-ECL-07902 & 49650.24560 & 0.00063 & Prim & B+R & MACHO \\
 OGLE-LMC-ECL-07902 & 49651.07718 & 0.00133 & Sec  & B+R & MACHO \\
 OGLE-LMC-ECL-07902 & 49999.80063 & 0.00101 & Prim & B+R & MACHO \\
 OGLE-LMC-ECL-07902 & 50000.63238 & 0.00467 & Sec  & B+R & MACHO \\
 \dots \\
  \noalign{\smallskip}\hline
\end{tabular}
(This table is available in its entirety in the appendix section at the end, or via the
CDS-tables.)
\end{minipage}
\end{table}



\begin{table*}[h!]
\caption{Identification of the analysed systems.}  \label{InfoSystems}
 \footnotesize
\begin{tabular}{lcccccccccccl}
   \hline\hline\noalign{\smallskip}
 System   & OGLE II$^{1}$ &  MACHO     &          RA           &             DE                              & $I_{\rm max}^{2}$ & $(V-I)_0^{3}$ & $(B-V)_0^{4}$  \\
  \hline\noalign{\smallskip}
 OGLE LMC-ECL-07902 & SC13 257494 & 19.4423.464  & 05$^h$07$^m$24$^s$.60 & -68$^\circ$29$^\prime$32$^{\prime\prime}$.4 & 16.69 &  -0.103 &  \\
 OGLE LMC-ECL-10133 & SC9 32388   & 79.5257.20   & 05$^h$13$^m$02$^s$.18 & -69$^\circ$19$^\prime$39$^{\prime\prime}$.8 & 15.60 &  -0.167 &  -0.154  \\
 OGLE LMC-ECL-10279 & SC9 115549  & 5.5376.2159  & 05$^h$13$^m$22$^s$.04 & -69$^\circ$27$^\prime$24$^{\prime\prime}$.9 & 16.99 &  -0.278 &  -0.145  \\
 OGLE LMC-ECL-12256 & SC7 120945  & 78.6096.420  & 05$^h$18$^m$15$^s$.55 & -69$^\circ$50$^\prime$54$^{\prime\prime}$.2 & 17.71 &         &  -0.095  \\
 OGLE LMC-ECL-13620 & SC6 323121  & 80.6708.5455 & 05$^h$21$^m$37$^s$.40 & -69$^\circ$24$^\prime$20$^{\prime\prime}$.9 & 17.83 &   0.023 &  -0.187$^{5}$ \\
 OGLE LMC-ECL-13666 & SC6 322419  & 78.6708.115  & 05$^h$21$^m$43$^s$.64 & -69$^\circ$24$^\prime$42$^{\prime\prime}$.1 & 16.21 &  -0.214 &  -0.185  \\
 OGLE LMC-ECL-14771 &             & 3.7084.72    & 05$^h$24$^m$15$^s$.96 & -68$^\circ$31$^\prime$20$^{\prime\prime}$.6 & 17.10 &  -0.209 &  -0.191  \\
 OGLE LMC-ECL-15895 & SC4 296290  & 77.7548.414  & 05$^h$26$^m$49$^s$.26 & -69$^\circ$49$^\prime$57$^{\prime\prime}$.2 & 16.93 &   0.120 &  -0.052  \\
 OGLE LMC-ECL-18102 &             & 82.8282.218  & 05$^h$31$^m$15$^s$.33 & -69$^\circ$20$^\prime$25$^{\prime\prime}$.0 & 17.24 &  -0.167 &  -0.107  \\
 OGLE LMC-ECL-19759 & SC16 70652  & 81.8881.44   & 05$^h$35$^m$02$^s$.02 & -69$^\circ$44$^\prime$17$^{\prime\prime}$.9 & 15.11 &  -0.360 &  -0.245$^{5}$ \\
 OGLE LMC-ECL-20112 &             & 81.9003.27   & 05$^h$35$^m$49$^s$.13 & -69$^\circ$37$^\prime$56$^{\prime\prime}$.6 & 14.78 &  -0.036 &  -0.226$^{5}$ \\
 OGLE LMC-ECL-20438 &             & 82.9130.43   & 05$^h$36$^m$27$^s$.18 & -69$^\circ$14$^\prime$15$^{\prime\prime}$.4 & 15.61 &  -0.109 &  -0.257  \\
 OGLE LMC-ECL-20498 &             & 82.9131.80   & 05$^h$36$^m$35$^s$.07 & -69$^\circ$10$^\prime$28$^{\prime\prime}$.2 & 16.15 &  -0.318 &  -0.265  \\
 \noalign{\smallskip}\hline
\end{tabular}
\\ 
 Notes: [1] - The full name from the OGLE II survey should be OGLE LMC-SCn nnnnnn, [2] -
Value taken from Graczyk et al. (2011), [3] - Value taken from Ulaczyk et al. (2012),
[4] - Value derived from the (B-V) and (U-B) indices taken from Massey (2002), [5] Value
taken from Zaritsky et al. (2004).
\end{table*}

\begin{table*}[h!]
\caption{Light curve parameters for the analysed systems.} \label{LCparam} \scriptsize
\begin{tabular}{lcccccccccc}
  \hline\hline\noalign{\smallskip}
   System   &  $T_1$ [K]   &  $T_2$ [K]  &  $i$ [deg]   & $q=M_2/M_1$ &  $\Omega_1$   &  $\Omega_2$   &  $L_1$ [\%]  &  $L_2$  [\%] & $L_3$ [\%] \\
  \hline\noalign{\smallskip}
 $\#$07902 & 20000 (fixed) & 14166 (172) & 83.28 (0.32) & 0.65 (0.02) & 4.466 (0.039) & 4.885 (0.061) & 77.4 (1.2)   & 20.7 (1.0)   &  1.9 (1.0) \\
 $\#$10133 & 16000 (fixed) & 18463 (295) & 77.75 (0.54) & 1.05 (0.10) & 5.989 (0.073) & 7.152 (0.096) & 53.2 (1.2)   & 44.8 (1.5)   &  2.0 (1.8) \\
 $\#$10279 & 15000 (fixed) & 14587 (106) & 87.57 (0.39) & 0.86 (0.06) & 4.661 (0.028) & 5.154 (0.037) & 62.2 (1.1)   & 37.8 (0.9)   &  0         \\
 $\#$12256 & 12000 (fixed) & 11768 (170) & 87.67 (0.70) & 1.08 (0.06) & 6.183 (0.130) & 5.669 (0.082) & 43.4 (1.6)   & 56.6 (1.5)   &  0         \\
 $\#$13620 & 18000 (fixed) & 14462 (421) & 83.22 (1.62) & 0.68 (0.10) & 4.795 (0.089) & 5.337 (0.136) & 61.7 (2.1)   & 20.5 (1.6)   & 17.8 (1.4) \\
 $\#$13666 & 18000 (fixed) & 19124 (193) & 84.39 (0.40) & 1.08 (0.05) & 5.868 (0.050) & 5.951 (0.046) & 45.2 (1.0)   & 52.8 (1.2)   &  2.0 (0.9) \\
 $\#$14771 & 18000 (fixed) & 13601 (283) & 83.87 (0.32) & 0.61 (0.11) & 6.358 (0.104) & 6.186 (0.112) & 72.1 (1.3)   & 25.9 (2.0)   &  2.0 (1.1) \\
 $\#$15895 & 10000 (fixed) &  8406 (175) & 89.79 (2.06) & 0.60 (0.13) & 4.042 (0.083) & 4.925 (0.076) & 35.7 (1.4)   &  7.9 (1.3)   & 56.4 (3.9) \\
 $\#$18102 & 12500 (fixed) & 11245 (139) & 88.18 (1.04) & 0.73 (0.07) & 6.052 (0.074) & 6.705 (0.063) & 70.7 (0.4)   & 29.3 (0.3)   &  0         \\
 $\#$19759 & 20000 (fixed) & 14424 (105) & 87.99 (0.32) & 0.50 (0.08) & 4.334 (0.021) & 4.792 (0.028) & 82.6 (0.5)   & 15.3 (0.6)   &  2.1 (0.4) \\
 $\#$20112 & 21000 (fixed) & 20199 (121) & 84.06 (0.96) & 0.78 (0.06) & 4.803 (0.035) & 5.649 (0.043) & 64.9 (1.9)   & 28.2 (1.6)   &  6.8 (3.6) \\
 $\#$20438 & 25000 (fixed) & 26172 (243) & 75.85 (0.48) & 1.00 (0.03) & 5.629 (0.057) & 6.206 (0.072) & 49.0 (1.2)   & 41.2 (0.9)   &  9.8 (1.9) \\
 $\#$20498 & 25000 (fixed) & 20373 (190) & 89.77 (0.32) & 0.63 (0.04) & 4.602 (0.026) & 4.848 (0.036) & 69.3 (0.7)   & 22.5 (0.3)   &  8.2 (0.8) \\
 \noalign{\smallskip}\hline
\end{tabular}
\end{table*}

\begin{table*}[h!]
\caption{The parameters of the apsidal motion for the individual systems.} \label{OCparam}
 \footnotesize
\begin{tabular}{lcccccccc}
\hline\hline\noalign{\smallskip}
  System   & $T_0 - 2400000$ [HJD] &  $P_s$ [days] &   $e$     & $\dot{\omega}$ [deg$/\rm{cycle}$] & $\omega_0$ [deg] & $U$ [yr]    \\ 
 \noalign{\smallskip}\hline\noalign{\smallskip}
 $\#$07902 & 52436.6397 (8)        & 1.6725077 (5) & 0.011  (3)& 0.0769 (110)                      & 286.2 (1.5)      &  21.4  (3.6)\\  
 $\#$10133 & 53566.7152 (31)       & 4.5023648 (52)& 0.194 (58)& 0.0413 (56)                       &  68.2 (5.4)      & 107.4 (16.8)\\  
 $\#$10279 & 53564.1812 (8)        & 1.7883577 (5) & 0.011 (3) & 0.0658 (83)                       & 185.3 (2.4)      &  26.8  (3.8)\\  
 $\#$12256 & 56701.5995 (115)      & 2.4137033 (53)& 0.105 (24)& 0.0283 (32)                       & 326.2 (3.3)      &  84.2 (10.6)\\  
 $\#$13620 & 53500.4749 (60)       & 2.1350145 (53)& 0.075 (20)& 0.0247 (70)                       & 321.4 (5.1)      &  85.1 (33.5)\\  
 $\#$13666 & 53503.1465 (188)      & 3.3881128 (27)& 0.124 (36)& 0.0332 (65)                       & 137.7 (2.9)      & 100.7 (24.4)\\  
 $\#$14771 & 53564.9091 (220)      & 2.1804949(188)& 0.295(172)& 0.0242 (60)                       & 273.7 (1.8)      &  88.8 (47.2)\\  
 $\#$15895 & 56652.5199 (239)      & 2.7098695(115)& 0.106 (25)& 0.0459 (99)                       &  65.5 (7.7)      &  58.1 (16.2)\\  
 $\#$18102 & 56945.0581 (425)      & 2.2650335(196)& 0.206 (64)& 0.0314 (47)                       & 106.5 (3.4)      &  71.1 (12.4)\\  
 $\#$19759 & 56974.3200 (252)      & 2.9895026 (14)& 0.102 (24)& 0.0459 (69)                       &  62.8 (7.5)      &  64.2 (11.4)\\  
 $\#$20112 & 56153.9786 (107)      & 3.0922295 (71)& 0.051 (12)& 0.0680 (35)                       & 219.8 (5.5)      &  44.8  (2.2)\\  
 $\#$20438 & 53572.6795 (131)      & 3.4134160 (28)& 0.070 (20)& 0.0422 (57)                       & 168.6 (8.2)      &  79.6 (12.4)\\  
 $\#$20498 & 53570.9957 (30)       & 2.0728518 (23)& 0.033 (9) & 0.0486 (80)                       & 348.7 (2.7)      &  42.0  (8.3)\\  
 \noalign{\smallskip}\hline
\end{tabular}
\end{table*}

\section{Notes on individual systems} \label{notes}

\begin{figure*}
 \includegraphics[width=\textwidth]{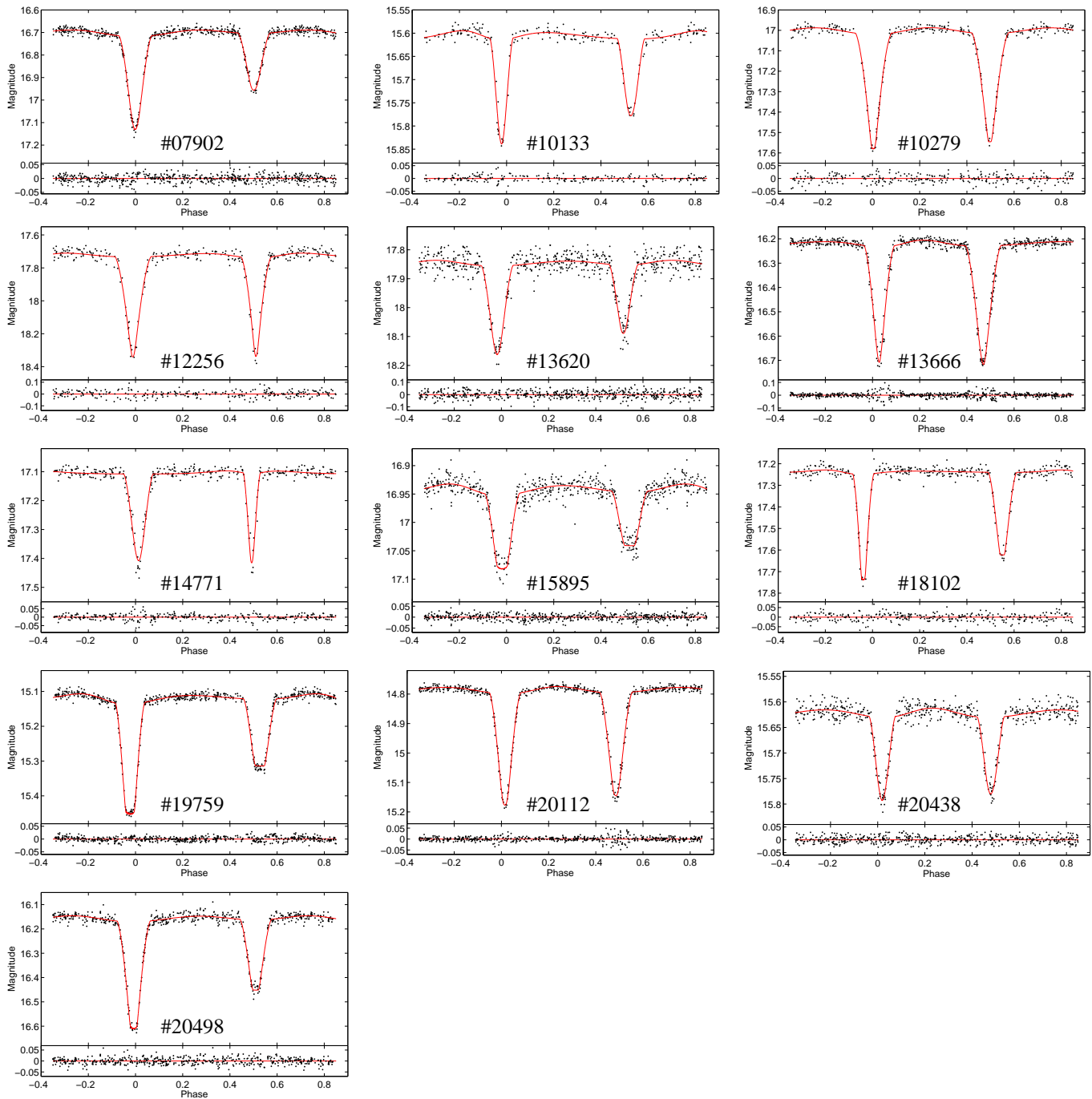}
 \caption[]{Light curves of the analysed systems, the data taken from the OGLE III survey, and the $I$
 filter. The bottom plots represent the residuals after subtraction of the fit.} \label{LCs}
\end{figure*}

\begin{figure*}
 \includegraphics[width=\textwidth]{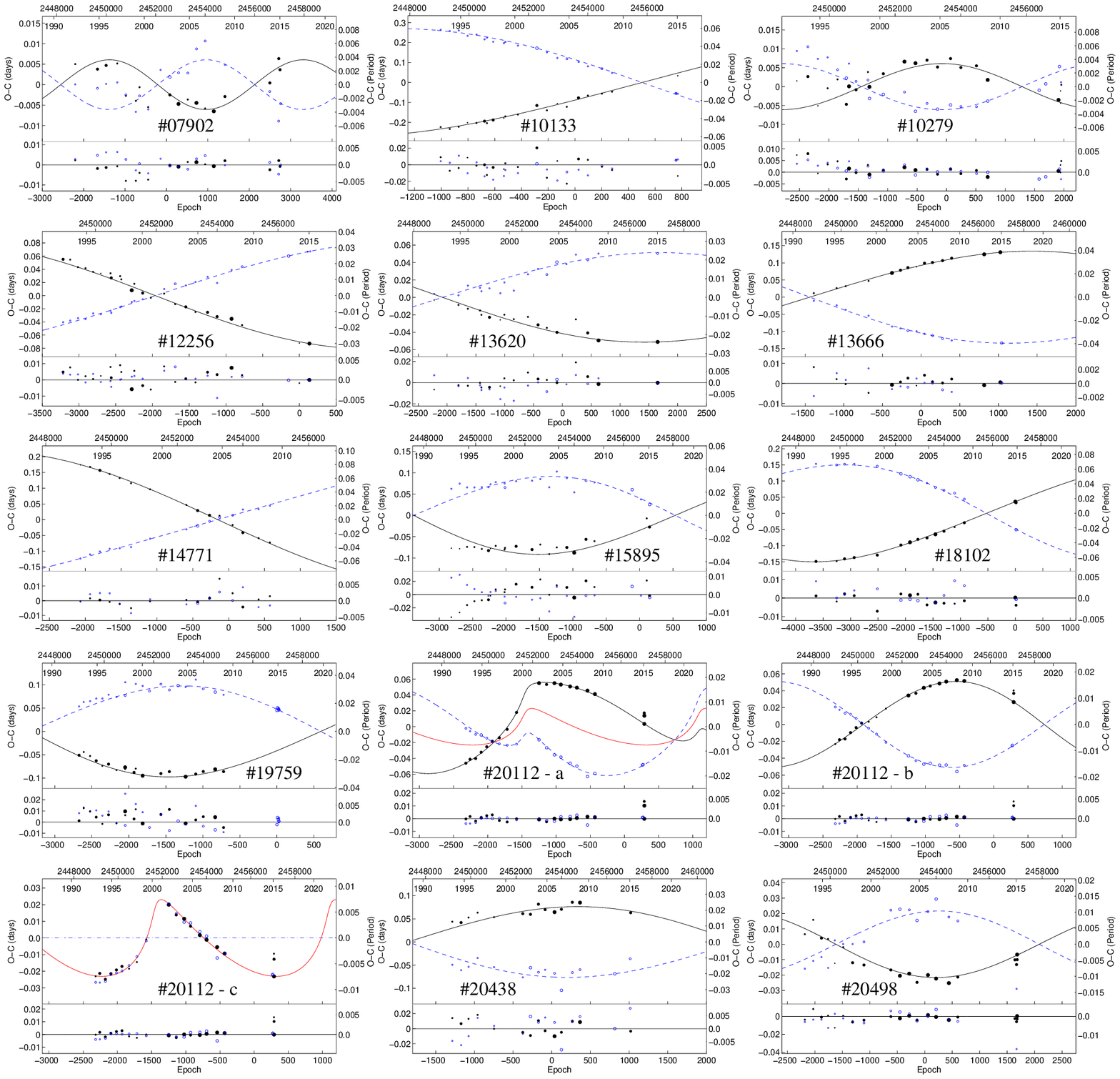}
 \caption[]{$O-C$ diagrams for the times of minima of the analysed systems. The continuous and dashed
curves represent predictions for the primary and secondary eclipses, respectively. The individual
primary and secondary minima are denoted by dots and open circles, respectively. Larger
symbols correspond to the more precise measurements. The bottom plots represent the residuals after
subtraction of the fit.} \label{OCs}
\end{figure*}

The eclipsing systems included in our analysis were proceeded in a similar way, hence we cannot
focus on every star in detail. For some information and cross-identification of the stars see Table
\ref{InfoSystems}. For abbreviating the star names we used the notation used for the {\sc Ogle III}
survey for a better brevity, hence e.g. OGLE-LMC-ECL-07902 was shortened as $\#$07902, etc. Only
the most important results are summarized below in a few subsections. The final light curve fits,
and the $O-C$ diagrams are presented in Figs. \ref{LCs} and \ref{OCs}; the parameters are given in
Tables \ref{LCparam} and \ref{OCparam}.

\subsection{OGLE LMC-ECL-07902}

For the only one system we found several spectroscopic data in the ESO archive. The star
OGLE~LMC-ECL-07902 was observed with the UVES (UV-Visual Echelle Spectrograph) during the ESO
period 68 and 70 programmes. The exposure times were from 3300 to 3600 seconds, while the typical
S/N ratios were from 12 to 35. All of the data were reduced using the standard ESO pipelines, and
the final radial velocities (hereafter RV) used for the analysis were derived via a manual
cross-correlation technique (i.e. the direct and flipped profile of the spectral lines manually
shifted on the computer screen to achieve the best match) using the program SPEFO (Horn et al.
1996; \v{S}koda 1996) on several absorption lines (usually HeI lines in the region from 370 to 420
nm). The derived radial velocities are given in Table \ref{RVs}, however their respective errors
are quite large due to weakness of some of the lines.

\begin{table}
\tiny \caption{List of the radial velocities used for the analysis.} \label{RVs} \scalebox{0.9}{
\begin{tabular}{cccccc}
\hline\hline\noalign{\smallskip}
 Star             &  JD Hel.- & $RV_1$      & error $RV_1$ & $RV_2$      &  error $RV_2$ \\
                  & 2400000   &[km s$^{-1}$]&[km s$^{-1}$] &[km s$^{-1}$]& [km s$^{-1}$] \\
\noalign{\smallskip}\hline \noalign{\smallskip}
 OGLE-LMC-ECL-07902 & 52249.27501 & 125.54 & 5.66 & 532.14 & 18.14 \\
 OGLE-LMC-ECL-07902 & 52249.31372 & 133.95 & 6.30 & 513.63 & 15.03 \\
 OGLE-LMC-ECL-07902 & 52250.22672 & 412.18 & 2.81 &  93.63 & 11.58 \\
 OGLE-LMC-ECL-07902 & 52250.18798 & 428.65 & 4.09 &  68.18 & 11.26 \\
 OGLE-LMC-ECL-07902 & 52622.04155 & 159.93 & 7.20 & 493.02 & 12.92 \\
\noalign{\smallskip}\hline
\end{tabular}}
\end{table}

\begin{table}[b] \caption{Radial velocity fit for $\#$07902.} \label{RV07902}
\begin{flushleft}
\begin{tabular}{lcccccc}
\hline\hline\noalign{\smallskip}
Parameter [Unit]  &   Value  \\
\noalign{\smallskip}\hline\noalign{\smallskip}
 $A$ [R$_\odot$]  & 13.26 $\pm$ 0.05 \\
 $q$ (=$M_2/M_1$) & 0.65 $\pm$ 0.02  \\ 
 $\gamma$ [km/s]  & 283.9 $\pm$ 0.4\\ \hline
 \multicolumn{2}{c}{Derived quantities:} \\
 $M_1 [M_\odot]$  & 6.8 $\pm$ 0.5 \\
 $M_2 [M_\odot]$  & 4.4 $\pm$ 0.4 \\
 $R_1 [R_\odot]$  & 3.5 $\pm$ 0.2 \\
 $R_2 [R_\odot]$  & 2.3 $\pm$ 0.2 \\
 \noalign{\smallskip}\hline
\end{tabular}
\end{flushleft}
\end{table}

\begin{figure}
 \includegraphics[width=0.48\textwidth]{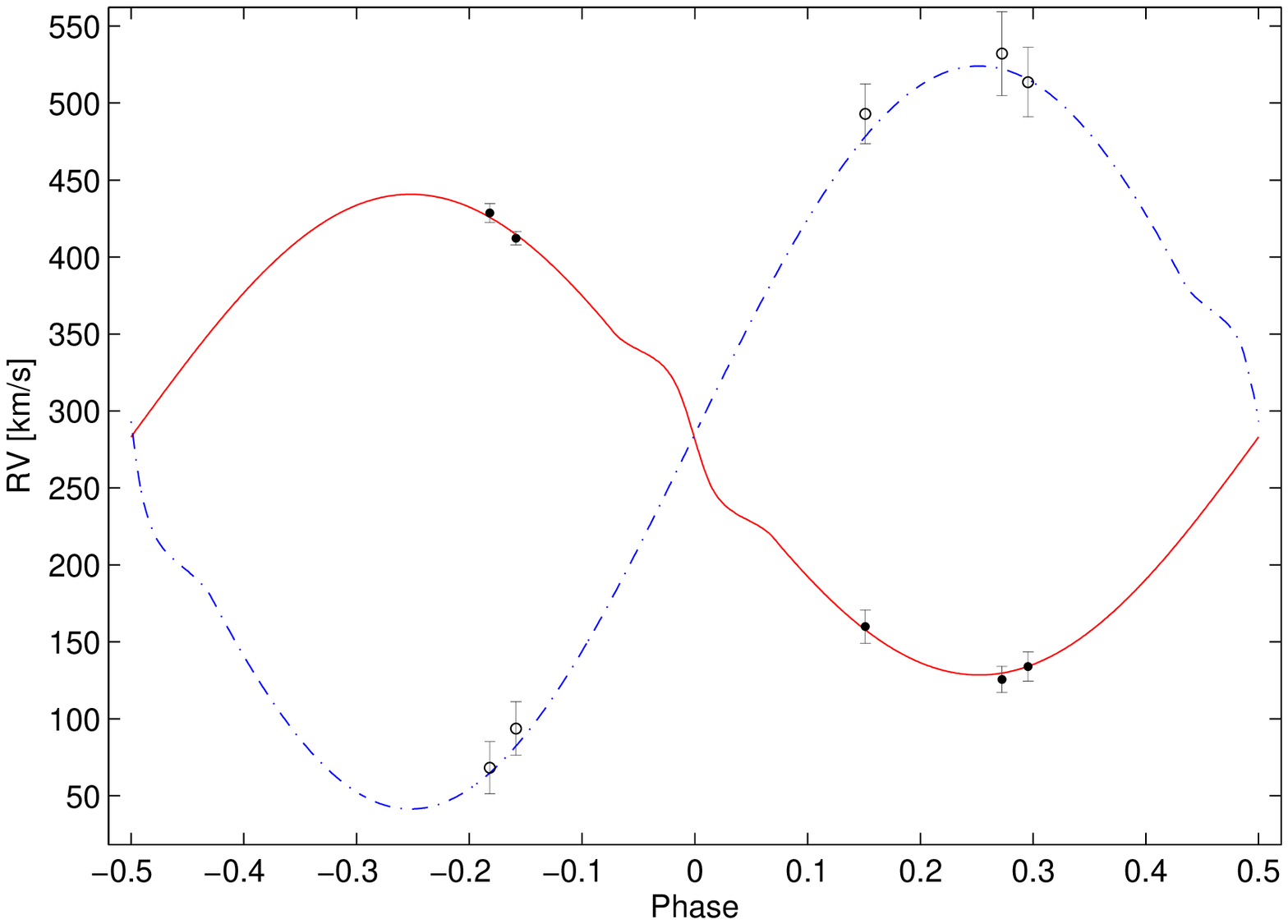}
 \caption[]{Radial velocity curves for the system OGLE LMC-ECL-07902 as derived from the ESO data,
 see the text for details.} \label{RV07902fig}
\end{figure}

We solved simultaneously the RVs with the LCs in our solution. The final RV curve fit is
presented in Fig. \ref{RV07902fig}, while the parameters are given in Table \ref{RV07902}. From
this combined solution there resulted that the eclipsing masses are of about 6.8 and 4.4~$M_\odot$,
which is in very good agreement with the assumed fixed primary temperature $T_1=20000$~K. As one
can see, the final fit provides us with a solution which is far from the original assumption that
the mass ratio $q=1$. However, for the other systems the spectroscopy is unavailable and
for deriving the mass ratio we have to use a method described above in section \ref{modelling}.
Moreover, there was also discovered that the system shows an emission behavior (in all of the
Balmer lines), but which remained fixed at a position of about +223 km$\cdot$s$^{-1}$. If such an
emission comes from the system or some other object in the same direction remains unsolved.

\medskip

\subsection{OGLE-LMC-ECL-20112}

Definitely the most interesting seems to be the object $\#$20112. We fixed the primary temperature
at a value of 21000~K in agreement with a rough spectral estimation of B2V according to Zaritsky et
al. (2004). With this temperature we found a LC solution (see Fig. \ref{LCs}), which was then used
as a template to derive the individual minima times for a subsequent period analysis.

Analysing the available times of minima observations we found that there is also an additional
variation after subtraction of the apsidal motion term in the $O-C$ diagram, hence we can speculate
that the system is probably a triple one. For the analysis we used a so-called 'light-travel time
effect', see e.g. Irwin (1959), or Mayer (1990), simultaneously with the apsidal motion.
The final plots with the fits are given in Fig.\ref{OCs}, where one can see the final complete fit
(noted as $\#$20112 - $a$), only the apsidal motion fit ($b$), and only the third-body fit ($c$).
The parameters are given in Table \ref{LITE20112}. Its 22-yr orbit is just being covered with the
individual observations yet, however some new observations would be of great benefit for the
hypothesis to be confirmed with higher conclusiveness.

\begin{table}[b]
\caption{Third-body orbit parameters for $\#$20112.} \label{LITE20112}
\begin{flushleft}
\begin{tabular}{lcccccc}
\hline\hline\noalign{\smallskip}
Parameter [Unit]     &   Value  \\
\noalign{\smallskip}\hline\noalign{\smallskip}
 $p_3$ [yr]        & 21.6 $\pm$ 7.1 \\
 $A_3$ [day]       & 0.0230 $\pm$ 0.0087  \\
 $T_3$ [HJD]       & 2459536 $\pm$ 1121  \\
 $e_3$             & 0.743 $\pm$ 0.088  \\
 $\omega_3$ [deg]  & 35.9 $\pm$ 9.0  \\
 \noalign{\smallskip}\hline
\end{tabular}
\end{flushleft}
\end{table}

From the third-body parameters, we are also able to compute the mass function of the distant
component, which resulted in $f(m_3) = 0.267 \pm 0.028$ \ms. From this value, one can calculate a
predicted minimal mass of the third body (i.e. assuming coplanar orbits $i_3=90^\circ$, and the
masses of the eclipsing components $M_1+M_2=14$~\ms), which resulted in $m_{3,min} = 4.5~$\ms. The
amplitude of radial velocity variation due to the third body resulted in about
7.7~km$\cdot$s$^{-1}$. If we propose such a body in the system, one can ask whether it is
detectable somehow in the already obtained data. The period is long for continuous monitoring of
the radial velocity changes, but detecting the third light in the light curve solution would be
promising. Assuming a normal main sequence star, its luminosity would be of about $L_{3,min}
\approx 10$\,\% of the total system luminosity. During our light curve fitting we detected a third
light of about 7\% of the total light, however its uncertainty is quite large.

\subsection{Systems with a third light}

For several systems the detected value of the third light is rather high, so we can speculate about
their triplicity. Of course, any such third light detection is not a direct evidence of a triple
star, because the contributing component does not have to be necessarily bounded to the eclipsing
system and can only be a so-called optical double. We detected a significant contribution of the
third light (dominating above an error, i.e. a limit of about 5\%) in 5 systems out of 13, i.e. in
about 38\% of the sample. However, even such a high number of the triple candidates cannot easily
be ruled out, because the fraction of triple systems for the stars of early spectral type is
generally high, see e.g. Pribulla \& Rucinski (2006), or Chini et al. (2012).

For $\#$15895 the 56\% contribution of the third light makes such a putative third component the
dominant star in the system and the depths of both eclipses are due to this reason only very
shallow. However, a variation in the $O-C$ diagram is still rather difficult to detect here. The
referee does not agree with the interpretation on system $\#$15895. According to him the light
curve of this system can be perfectly fitted without the third light for the mass ratio
q~$\approx$~0.3. However, we were not able to reproduce such a result with the lowest rms.

\subsection{Other interesting systems}

Some of the systems seem to show rather high scatter of residuals after subtracting the
apsidal motion term. However, it cannot easily be described by a single periodic modulation or our
data are too sparse and the time span is still rather short. Such systems are e.g. $\#$07902,
$\#$10279, $\#$13620, $\#$14771, $\#$20438, and $\#$20498. The residuals plotted in Figure
\ref{OCs} show that the poor coverage of the data cannot allow us to derive any period or amplitude
of such prospective modulation. Only more updated observations for these systems can help us to
prove or reject any such hypothesis.

\section{Discussion and conclusions}

We performed the first analysis of the apsidal motion and the LC fitting for thirteen early-type
binary systems from the Large Magellanic Cloud. In our own Galaxy there are a few hundreds of
apsidal motion eclipsing binaries known (Bulut \& Demircan 2007); however, in other galaxies their
number is still only very limited. Hence, this study still presents an important contribution to
this topic. For some of the systems the presented apsidal motion hypothesis is still rather
preliminary yet due to poor coverage of the $O-C$ diagrams with the observations. For some others
the fits as presented in Fig.\ref{OCs} are fairly reliable with the current data set.

For the system $\#$20112 we presented a third-body hypothesis, which resulted from the analysis of
residuals after subtracting the apsidal motion term in the $O-C$ diagram. Its 22-yr variation is
still preliminary, and should be confirmed via dedicated new observations in upcoming years. We
also presented a few more similar systems, which are suspicious to be also triples, but for which
much more observations are needed.

There also resulted that the presented apsidal and light curve analysis without any spectroscopic
information cannot be used for testing the evolutionary stellar models via deriving the internal
structure constant values. Their errors are too high and a more thorough analysis with some radial
velocities would be needed. Some of the presented stars are bright enough for a spectral
monitoring, hence we encourage the observers to obtain new, high-dispersion, and high-S/N
spectroscopic observations. Using these data and methods like spectral disentangling can help us to
construct the radial velocity curves of both components, confirm the apsidal motion hypothesis,
test the stellar structure models, or detect the third bodies, as indicated from our analysis. The
same apply for testing the models for slightly different chemical composition of LMC stars (e.g.
Ribas 2004). Much better data are needed, however our presented analysis can serve as a starting
point for these dedicated observations.

\medskip

\begin{acknowledgements}
We do thank the {\sc MACHO} and {\sc OGLE} teams for making all of the observations easily public
available. This work was supported by the Czech Science Foundation grants no. P209/10/0715, and
GA15-02112S, and also by the student's project SVV-260089. We are also grateful to the ESO team at
the La Silla Observatory for their help in maintaining and operating the Danish telescope. The
following internet-based resources were used in research for this paper: the SIMBAD database and
the VizieR service operated at the CDS, Strasbourg, France, and the NASA's Astrophysics Data System
Bibliographic Services.
\end{acknowledgements}


\begin{appendix} 

\section{Tables of minima}

\begin{table}[b]
 \centering
  \begin{minipage}{95mm}
 \fontsize{1.8mm}{2.4mm}\selectfont
 \caption{List of the minima timings used for the analysis.} \label{minima}
\begin{tabular}{ccclcl}
\hline\hline\noalign{\smallskip}
 Star       &    JD Hel.- &  Error & Type   &  Filter  & Source /     \\
            &   2400000   &  [day] &        &          & Observatory  \\
\noalign{\smallskip}\hline
\noalign{\smallskip}
 OGLE-LMC-ECL-07902 & 48750.43767 & 0.00180 & Prim & B+R & MACHO \\
 OGLE-LMC-ECL-07902 & 48751.26899 & 0.00262 & Sec  & B+R & MACHO \\
 OGLE-LMC-ECL-07902 & 49650.24560 & 0.00063 & Prim & B+R & MACHO \\
 OGLE-LMC-ECL-07902 & 49651.07718 & 0.00133 & Sec  & B+R & MACHO \\
 OGLE-LMC-ECL-07902 & 49999.80063 & 0.00101 & Prim & B+R & MACHO \\
 OGLE-LMC-ECL-07902 & 50000.63238 & 0.00467 & Sec  & B+R & MACHO \\
 OGLE-LMC-ECL-07902 & 50449.70569 & 0.00191 & Prim & B+R & MACHO \\
 OGLE-LMC-ECL-07902 & 50450.53731 & 0.00478 & Sec  & B+R & MACHO \\
 OGLE-LMC-ECL-07902 & 51349.50906 & 0.00397 & Prim & B+R & MACHO \\
 OGLE-LMC-ECL-07902 & 51350.34318 & 0.00133 & Sec  & B+R & MACHO \\
 OGLE-LMC-ECL-07902 & 50799.25190 & 0.00207 & Prim &  I  & OGLE II  \\
 OGLE-LMC-ECL-07902 & 50800.08843 & 0.00129 & Sec  &  I  & OGLE II  \\
 OGLE-LMC-ECL-07902 & 51200.65253 & 0.00173 & Prim &  I  & OGLE II  \\
 OGLE-LMC-ECL-07902 & 51201.48611 & 0.00244 & Sec  &  I  & OGLE II  \\
 OGLE-LMC-ECL-07902 & 51700.73081 & 0.00174 & Prim &  I  & OGLE II  \\
 OGLE-LMC-ECL-07902 & 51701.56651 & 0.00321 & Sec  &  I  & OGLE II  \\
 OGLE-LMC-ECL-07902 & 52201.65596 & 0.00423 & Sec  &  I  & OGLE III \\
 OGLE-LMC-ECL-07902 & 52575.45535 & 0.00071 & Prim &  I  & OGLE III \\
 OGLE-LMC-ECL-07902 & 52576.29597 & 0.00034 & Sec  &  I  & OGLE III \\
 OGLE-LMC-ECL-07902 & 52925.00725 & 0.00046 & Prim &  I  & OGLE III \\
 OGLE-LMC-ECL-07902 & 52925.85112 & 0.00144 & Sec  &  I  & OGLE III \\
 OGLE-LMC-ECL-07902 & 53299.65006 & 0.00155 & Prim &  I  & OGLE III \\
 OGLE-LMC-ECL-07902 & 53300.49278 & 0.00129 & Sec  &  I  & OGLE III \\
 OGLE-LMC-ECL-07902 & 53649.20336 & 0.00032 & Prim &  I  & OGLE III \\
 OGLE-LMC-ECL-07902 & 53650.05280 & 0.00070 & Sec  &  I  & OGLE III \\
 OGLE-LMC-ECL-07902 & 54000.42863 & 0.00201 & Prim &  I  & OGLE III \\
 OGLE-LMC-ECL-07902 & 54001.28134 & 0.00060 & Sec  &  I  & OGLE III \\
 OGLE-LMC-ECL-07902 & 54349.98205 & 0.00030 & Prim &  I  & OGLE III \\
 OGLE-LMC-ECL-07902 & 54799.89023 & 0.00149 & Prim &  I  & OGLE III \\
 OGLE-LMC-ECL-07902 & 54800.73423 & 0.00157 & Sec  &  I  & OGLE III \\
 OGLE-LMC-ECL-07902 & 56608.70858 & 0.00099 & Sec  &  R  & DK154 \\
 OGLE-LMC-ECL-07902 & 56614.56445 & 0.00079 & Prim &  R  & DK154 \\
 OGLE-LMC-ECL-07902 & 56971.63552 & 0.00085 & Sec  &  R  & DK154 \\
 OGLE-LMC-ECL-07902 & 56975.83205 & 0.00119 & Prim &  R  & DK154 \\
 OGLE-LMC-ECL-07902 & 57027.67709 & 0.00090 & Prim &  R  & DK154 \\
 OGLE-LMC-ECL-07902 & 57053.59265 & 0.00129 & Sec  &  R  & DK154 \\  \hline
 OGLE-LMC-ECL-10133 & 52283.42750 &   0.00396 &   Prim  &    I    &  OGLE III \\
 OGLE-LMC-ECL-10133 & 52285.96599 &   0.00054 &   Sec   &    I    &  OGLE III \\
 OGLE-LMC-ECL-10133 & 52621.08091 &   0.00697 &   Prim  &    I    &  OGLE III \\
 OGLE-LMC-ECL-10133 & 52623.62263 &   0.00425 &   Sec   &    I    &  OGLE III \\
 OGLE-LMC-ECL-10133 & 52972.29867 &   0.01157 &   Prim  &    I    &  OGLE III \\
 OGLE-LMC-ECL-10133 & 52974.77879 &   0.00428 &   Sec   &    I    &  OGLE III \\
 OGLE-LMC-ECL-10133 & 53291.95111 &   0.00448 &   Prim  &    I    &  OGLE III \\
 OGLE-LMC-ECL-10133 & 53294.44267 &   0.00352 &   Sec   &    I    &  OGLE III \\
 OGLE-LMC-ECL-10133 & 53688.20416 &   0.00395 &   Prim  &    I    &  OGLE III \\
 OGLE-LMC-ECL-10133 & 53690.63503 &   0.00659 &   Sec   &    I    &  OGLE III \\
 OGLE-LMC-ECL-10133 & 54003.38087 &   0.00951 &   Prim  &    I    &  OGLE III \\
 OGLE-LMC-ECL-10133 & 54005.77866 &   0.00463 &   Sec   &    I    &  OGLE III \\
 OGLE-LMC-ECL-10133 & 54476.13926 &   0.00598 &   Prim  &    I    &  OGLE III \\
 OGLE-LMC-ECL-10133 & 54478.51529 &   0.00173 &   Sec   &    I    &  OGLE III \\
 OGLE-LMC-ECL-10133 & 54818.32862 &   0.00928 &   Prim  &    I    &  OGLE III \\
 OGLE-LMC-ECL-10133 & 54820.67411 &   0.00462 &   Sec   &    I    &  OGLE III \\
 OGLE-LMC-ECL-10133 & 50518.42151 &   0.00202 &   Prim  &    I    &  OGLE II  \\
 OGLE-LMC-ECL-10133 & 50521.08229 &   0.00589 &   Sec   &    I    &  OGLE II  \\
 OGLE-LMC-ECL-10133 & 50811.08100 &   0.00343 &   Prim  &    I    &  OGLE II \\
 OGLE-LMC-ECL-10133 & 50813.72398 &   0.00796 &   Sec   &    I    &  OGLE II \\
 OGLE-LMC-ECL-10133 & 51220.81424 &   0.00545 &   Prim  &    I    &  OGLE II \\
 OGLE-LMC-ECL-10133 & 51223.43846 &   0.01620 &   Sec   &    I    &  OGLE II \\
 OGLE-LMC-ECL-10133 & 51644.04625 &   0.00432 &   Prim  &    I    &  OGLE II \\
 OGLE-LMC-ECL-10133 & 51646.64283 &   0.00906 &   Sec   &    I    &  OGLE II \\
 OGLE-LMC-ECL-10133 & 49041.61566 &   0.00650 &   Prim  &    B+R  &  MACHO \\
 OGLE-LMC-ECL-10133 & 49044.35420 &   0.01014 &   Sec   &    B+R  &  MACHO \\
 OGLE-LMC-ECL-10133 & 49347.76725 &   0.00927 &   Prim  &    B+R  &  MACHO \\
 OGLE-LMC-ECL-10133 & 49350.50869 &   0.00921 &   Sec   &    B+R  &  MACHO \\
 OGLE-LMC-ECL-10133 & 49649.43670 &   0.00738 &   Prim  &    B+R  &  MACHO \\
 OGLE-LMC-ECL-10133 & 49652.17045 &   0.01600 &   Sec   &    B+R  &  MACHO \\
 OGLE-LMC-ECL-10133 & 49946.61326 &   0.01300 &   Prim  &    B+R  &  MACHO \\
 OGLE-LMC-ECL-10133 & 49949.29942 &   0.00782 &   Sec   &    B+R  &  MACHO \\
 OGLE-LMC-ECL-10133 & 50248.27053 &   0.00676 &   Prim  &    B+R  &  MACHO \\
 OGLE-LMC-ECL-10133 & 50250.96420 &   0.00928 &   Sec   &    B+R  &  MACHO \\
 OGLE-LMC-ECL-10133 & 50599.45414 &   0.00979 &   Prim  &    B+R  &  MACHO \\
 OGLE-LMC-ECL-10133 & 50602.13333 &   0.02528 &   Sec   &    B+R  &  MACHO \\
 OGLE-LMC-ECL-10133 & 51171.27955 &   0.01578 &   Prim  &    B+R  &  MACHO \\
 OGLE-LMC-ECL-10133 & 51173.90678 &   0.00681 &   Sec   &    B+R  &  MACHO \\
 OGLE-LMC-ECL-10133 & 56972.70306 &   0.00315 &   Sec   &    R    &  DK154 \\
 OGLE-LMC-ECL-10133 & 56972.70371 &   0.00197 &   Sec   &    R    &  DK154 \\
 OGLE-LMC-ECL-10133 & 57033.57280 &   0.00443 &   Prim  &    R    &  DK154 \\
 OGLE-LMC-ECL-10133 & 57035.73398 &   0.00312 &   Sec   &    R    &  DK154 \\
 OGLE-LMC-ECL-10279 &   52285.51202 &   0.00043 &   Prim  &    I    &  OGLE III  \\
 OGLE-LMC-ECL-10279 &   52286.39815 &   0.00126 &   Sec   &    I    &  OGLE III  \\
  \noalign{\smallskip}\hline
\end{tabular}
\end{minipage}
\end{table}

\begin{table}[b]
 \centering
  \begin{minipage}{95mm}
 \fontsize{1.8mm}{2.4mm}\selectfont
 \caption{List of the minima timings used for the analysis - cont.}
\begin{tabular}{ccclcl}
\hline\hline\noalign{\smallskip}
 Star       &    JD Hel.- &  Error & Type   &  Filter  & Source /     \\
            &   2400000   &  [day] &        &          & Observatory  \\
\noalign{\smallskip}\hline
\noalign{\smallskip}
 OGLE-LMC-ECL-10279 &   52619.93450 &   0.00073 &   Prim  &    I    &  OGLE III  \\
 OGLE-LMC-ECL-10279 &   52620.81597 &   0.00056 &   Sec   &    I    &  OGLE III  \\
 OGLE-LMC-ECL-10279 &   52970.45341 &   0.00113 &   Prim  &    I    &  OGLE III  \\
 OGLE-LMC-ECL-10279 &   52971.33617 &   0.00168 &   Sec   &    I    &  OGLE III  \\
 OGLE-LMC-ECL-10279 &   53290.56763 &   0.00107 &   Prim  &    I    &  OGLE III  \\
 OGLE-LMC-ECL-10279 &   53291.45068 &   0.00091 &   Sec   &    I    &  OGLE III  \\
 OGLE-LMC-ECL-10279 &   53687.58529 &   0.00120 &   Prim  &    I    &  OGLE III  \\
 OGLE-LMC-ECL-10279 &   53688.46721 &   0.00048 &   Sec   &    I    &  OGLE III  \\
 OGLE-LMC-ECL-10279 &   54002.33388 &   0.00109 &   Prim  &    I    &  OGLE III  \\
 OGLE-LMC-ECL-10279 &   54003.21775 &   0.00204 &   Sec   &    I    &  OGLE III  \\
 OGLE-LMC-ECL-10279 &   54476.24913 &   0.00107 &   Prim  &    I    &  OGLE III  \\
 OGLE-LMC-ECL-10279 &   54477.13249 &   0.00173 &   Sec   &    I    &  OGLE III  \\
 OGLE-LMC-ECL-10279 &   54817.82173 &   0.00062 &   Prim  &    I    &  OGLE III  \\
 OGLE-LMC-ECL-10279 &   54818.71031 &   0.00050 &   Sec   &    I    &  OGLE III  \\
 OGLE-LMC-ECL-10279 &   50516.81496 &   0.00124 &   Prim  &    I    &  OGLE II   \\
 OGLE-LMC-ECL-10279 &   50517.71608 &   0.00078 &   Sec   &    I    &  OGLE II   \\
 OGLE-LMC-ECL-10279 &   50811.89782 &   0.00140 &   Prim  &    I    &  OGLE II \\
 OGLE-LMC-ECL-10279 &   50812.79423 &   0.00179 &   Sec   &    I    &  OGLE II \\
 OGLE-LMC-ECL-10279 &   51219.64415 &   0.00025 &   Prim  &    I    &  OGLE II \\
 OGLE-LMC-ECL-10279 &   51220.53527 &   0.00054 &   Sec   &    I    &  OGLE II \\
 OGLE-LMC-ECL-10279 &   51643.48841 &   0.00116 &   Prim  &    I    &  OGLE II \\
 OGLE-LMC-ECL-10279 &   51644.37713 &   0.00075 &   Sec   &    I    &  OGLE II \\
 OGLE-LMC-ECL-10279 &   49011.02394 &   0.00790 &   Prim  &    B+R  &  MACHO   \\
 OGLE-LMC-ECL-10279 &   49011.92602 &   0.00253 &   Sec   &    B+R  &  MACHO   \\
 OGLE-LMC-ECL-10279 &   49349.02470 &   0.00154 &   Prim  &    B+R  &  MACHO   \\
 OGLE-LMC-ECL-10279 &   49349.92675 &   0.00204 &   Sec   &    B+R  &  MACHO   \\
 OGLE-LMC-ECL-10279 &   49651.24942 &   0.00750 &   Prim  &    B+R  &  MACHO   \\
 OGLE-LMC-ECL-10279 &   49652.15602 &   0.00595 &   Sec   &    B+R  &  MACHO   \\
 OGLE-LMC-ECL-10279 &   49948.12145 &   0.00286 &   Prim  &    B+R  &  MACHO   \\
 OGLE-LMC-ECL-10279 &   49949.02378 &   0.00996 &   Sec   &    B+R  &  MACHO   \\
 OGLE-LMC-ECL-10279 &   50248.56788 &   0.00317 &   Prim  &    B+R  &  MACHO   \\
 OGLE-LMC-ECL-10279 &   50249.46461 &   0.00211 &   Sec   &    B+R  &  MACHO   \\
 OGLE-LMC-ECL-10279 &   50599.08417 &   0.00084 &   Prim  &    B+R  &  MACHO   \\
 OGLE-LMC-ECL-10279 &   50599.98436 &   0.01153 &   Sec   &    B+R  &  MACHO   \\
 OGLE-LMC-ECL-10279 &   51174.93913 &   0.00297 &   Prim  &    B+R  &  MACHO     \\
 OGLE-LMC-ECL-10279 &   51175.83120 &   0.00192 &   Sec   &    B+R  &  MACHO     \\
 OGLE-LMC-ECL-10279 &   56383.52618 &   0.00085 &   Sec   &    R    &  DK154     \\
 OGLE-LMC-ECL-10279 &   56583.82393 &   0.00000 &   Sec   &    R    &  DK154     \\
 OGLE-LMC-ECL-10279 &   56972.78749 &   0.00047 &   Prim  &    R    &  DK154     \\
 OGLE-LMC-ECL-10279 &   56998.72750 &   0.00055 &   Sec   &    R    &  DK154     \\
 OGLE-LMC-ECL-10279 &   57031.80569 &   0.02887 &   Prim  &    R    &  DK154     \\
 OGLE-LMC-ECL-10279 &   57033.59083 &   0.00842 &   Prim  &    R    &  DK154     \\
 OGLE-LMC-ECL-10279 &   57050.58574 &   0.00436 &   Sec   &    R    &  DK154     \\
 OGLE-LMC-ECL-10279 &   57056.84431 &   0.00590 &   Prim  &    R    &  DK154     \\  \hline
 OGLE-LMC-ECL-12256 &   52257.97404 &   0.00552 &   Prim  &    I    &  OGLE III  \\
 OGLE-LMC-ECL-12256 &   52259.18778 &   0.00319 &   Sec   &    I    &  OGLE III  \\
 OGLE-LMC-ECL-12256 &   52622.42726 &   0.00337 &   Prim  &    I    &  OGLE III  \\
 OGLE-LMC-ECL-12256 &   52623.66506 &   0.00159 &   Sec   &    I    &  OGLE III  \\
 OGLE-LMC-ECL-12256 &   52969.99645 &   0.00095 &   Prim  &    I    &  OGLE III  \\
 OGLE-LMC-ECL-12256 &   52971.23588 &   0.00278 &   Sec   &    I    &  OGLE III  \\
 OGLE-LMC-ECL-12256 &   53288.59761 &   0.00454 &   Prim  &    I    &  OGLE III  \\
 OGLE-LMC-ECL-12256 &   53289.85007 &   0.00460 &   Sec   &    I    &  OGLE III  \\
 OGLE-LMC-ECL-12256 &   53684.44457 &   0.00243 &   Prim  &    I    &  OGLE III  \\
 OGLE-LMC-ECL-12256 &   53685.70515 &   0.00356 &   Sec   &    I    &  OGLE III  \\
 OGLE-LMC-ECL-12256 &   53998.21947 &   0.00209 &   Prim  &    I    &  OGLE III  \\
 OGLE-LMC-ECL-12256 &   53999.47755 &   0.00543 &   Sec   &    I    &  OGLE III  \\
 OGLE-LMC-ECL-12256 &   54473.71585 &   0.00080 &   Prim  &    I    &  OGLE III  \\
 OGLE-LMC-ECL-12256 &   54474.99689 &   0.00355 &   Sec   &    I    &  OGLE III  \\
 OGLE-LMC-ECL-12256 &   54818.86588 &   0.00258 &   Prim  &    I    &  OGLE III  \\
 OGLE-LMC-ECL-12256 &   54820.16145 &   0.00189 &   Sec   &    I    &  OGLE III  \\
 OGLE-LMC-ECL-12256 &   50520.13102 &   0.00211 &   Prim  &    I    &  OGLE II   \\
 OGLE-LMC-ECL-12256 &   50521.28478 &   0.00308 &   Sec   &    I    &  OGLE II   \\
 OGLE-LMC-ECL-12256 &   50812.19189 &   0.00401 &   Prim  &    I    &  OGLE II   \\
 OGLE-LMC-ECL-12256 &   50813.35170 &   0.00335 &   Sec   &    I    &  OGLE II   \\
 OGLE-LMC-ECL-12256 &   51193.53566 &   0.00072 &   Prim  &    I    &  OGLE II   \\
 OGLE-LMC-ECL-12256 &   51194.72450 &   0.00284 &   Sec   &    I    &  OGLE II   \\
 OGLE-LMC-ECL-12256 &   51553.17333 &   0.00157 &   Prim  &    I    &  OGLE II   \\
 OGLE-LMC-ECL-12256 &   51554.37058 &   0.00293 &   Sec   &    I    &  OGLE II   \\
 OGLE-LMC-ECL-12256 &   51821.08724 &   0.00443 &   Prim  &    I    &  OGLE II   \\
 OGLE-LMC-ECL-12256 &   51554.37058 &   0.00293 &   Sec   &    I    &  OGLE II   \\
 OGLE-LMC-ECL-12256 &   48939.18328 &   0.00229 &   Prim  &    B+R  &  MACHO     \\
 OGLE-LMC-ECL-12256 &   48940.29459 &   0.00429 &   Sec   &    B+R  &  MACHO     \\
 OGLE-LMC-ECL-12256 &   49170.89838 &   0.00428 &   Prim  &    B+R  &  MACHO     \\
 OGLE-LMC-ECL-12256 &   49172.01264 &   0.00754 &   Sec   &    B+R  &  MACHO     \\
 OGLE-LMC-ECL-12256 &   49426.73954 &   0.00418 &   Prim  &    B+R  &  MACHO     \\
 OGLE-LMC-ECL-12256 &   49427.86887 &   0.00296 &   Sec   &    B+R &   MACHO     \\
 OGLE-LMC-ECL-12256 &   49675.34946 &   0.00561 &   Prim  &    B+R &   MACHO     \\
 OGLE-LMC-ECL-12256 &   49676.47890 &   0.00871 &   Sec   &    B+R &   MACHO     \\
 OGLE-LMC-ECL-12256 &   49923.95533 &   0.00445 &   Prim  &    B+R &   MACHO     \\
 OGLE-LMC-ECL-12256 &   49925.09917 &   0.00847 &   Sec   &    B+R &   MACHO     \\
  \noalign{\smallskip}\hline
\end{tabular}
\end{minipage}
\end{table}

\begin{table}[b]
 \centering
  \begin{minipage}{95mm}
 \fontsize{1.8mm}{2.4mm}\selectfont
 \caption{List of the minima timings used for the analysis - cont.}
\begin{tabular}{ccclcl}
\hline\hline\noalign{\smallskip}
 Star       &    JD Hel.- &  Error & Type   &  Filter  & Source /     \\
            &   2400000   &  [day] &        &          & Observatory  \\
\noalign{\smallskip}\hline
\noalign{\smallskip}
 OGLE-LMC-ECL-12256 &   50177.39195 &   0.00802 &   Prim  &    B+R  &  MACHO    \\
 OGLE-LMC-ECL-12256 &   50178.53678 &   0.00466 &   Sec   &    B+R  &  MACHO    \\
 OGLE-LMC-ECL-12256 &   50496.00101 &   0.00661 &   Prim  &    B+R  &  MACHO    \\
 OGLE-LMC-ECL-12256 &   50497.15088 &   0.00429 &   Sec   &    B+R  &  MACHO    \\
 OGLE-LMC-ECL-12256 &   50850.80633 &   0.00673 &   Prim  &    B+R  &  MACHO    \\
 OGLE-LMC-ECL-12256 &   50851.97181 &   0.00872 &   Sec   &    B+R  &  MACHO    \\
 OGLE-LMC-ECL-12256 &   51278.02511 &   0.00305 &   Prim  &    B+R  &  MACHO    \\
 OGLE-LMC-ECL-12256 &   51279.20583 &   0.00481 &   Sec   &    B+R  &  MACHO    \\
 OGLE-LMC-ECL-12256 &   56345.63853 &   0.00079 &   Sec   &    R    &  DK154    \\
 OGLE-LMC-ECL-12256 &   56701.52767 &   0.01220 &   Prim  &    R    &  DK154    \\
 OGLE-LMC-ECL-12256 &   56975.62188 &   0.01757 &   Sec   &    R    &  DK154     \\
 OGLE-LMC-ECL-12256 &   57029.79059 &   0.00015 &   Prim  &    R    &  DK154     \\
 OGLE-LMC-ECL-12256 &   57057.68825 &   0.00588 &   Sec   &    R    &  DK154     \\  \hline
 OGLE-LMC-ECL-13620 &   48750.06458 &   0.00716 &   Prim  &    B+R  &  MACHO     \\
 OGLE-LMC-ECL-13620 &   48751.13908 &   0.01388 &   Sec   &    B+R  &  MACHO     \\
 OGLE-LMC-ECL-13620 &   49651.03505 &   0.00520 &   Prim  &    B+R  &  MACHO     \\
 OGLE-LMC-ECL-13620 &   49652.11366 &   0.00387 &   Sec   &    B+R  &  MACHO     \\
 OGLE-LMC-ECL-13620 &   49999.03806 &   0.00474 &   Prim  &    B+R  &  MACHO     \\
 OGLE-LMC-ECL-13620 &   50000.13212 &   0.00983 &   Sec   &    B+R  &  MACHO     \\
 OGLE-LMC-ECL-13620 &   50449.52412 &   0.00647 &   Prim  &    B+R  &  MACHO     \\
 OGLE-LMC-ECL-13620 &   50450.61769 &   0.00695 &   Sec   &    B+R  &  MACHO     \\
 OGLE-LMC-ECL-13620 &   51350.49545 &   0.00653 &   Prim  &    B+R  &  MACHO     \\
 OGLE-LMC-ECL-13620 &   51351.60003 &   0.00846 &   Sec   &    B+R  &  MACHO     \\
 OGLE-LMC-ECL-13620 &   50500.75955 &   0.00613 &   Prim  &    I    &  OGLE II   \\
 OGLE-LMC-ECL-13620 &   50501.85405 &   0.00498 &   Sec   &    I    &  OGLE II   \\
 OGLE-LMC-ECL-13620 &   50799.65873 &   0.00203 &   Prim  &    I    &  OGLE II   \\
 OGLE-LMC-ECL-13620 &   50800.75949 &   0.00579 &   Sec   &    I    &  OGLE II   \\
 OGLE-LMC-ECL-13620 &   51201.03850 &   0.00895 &   Prim  &    I    &  OGLE II   \\
 OGLE-LMC-ECL-13620 &   51202.13661 &   0.01010 &   Sec   &    I    &  OGLE II   \\
 OGLE-LMC-ECL-13620 &   51700.63267 &   0.00630 &   Prim  &    I    &  OGLE II   \\
 OGLE-LMC-ECL-13620 &   51701.73293 &   0.01224 &   Sec   &    I    &  OGLE II   \\
 OGLE-LMC-ECL-13620 &   52225.84891 &   0.00395 &   Prim  &    I    &  OGLE III  \\
 OGLE-LMC-ECL-13620 &   52226.96530 &   0.00561 &   Sec   &    I    &  OGLE III  \\
 OGLE-LMC-ECL-13620 &   52599.46726 &   0.00174 &   Prim  &    I    &  OGLE III  \\
 OGLE-LMC-ECL-13620 &   52600.59886 &   0.00637 &   Sec   &    I    &  OGLE III  \\
 OGLE-LMC-ECL-13620 &   52923.98609 &   0.00258 &   Prim  &    I    &  OGLE III  \\
 OGLE-LMC-ECL-13620 &   52925.11474 &   0.00208 &   Sec   &    I    &  OGLE III  \\
 OGLE-LMC-ECL-13620 &   53299.74348 &   0.00096 &   Prim  &    I    &  OGLE III  \\
 OGLE-LMC-ECL-13620 &   53300.89148 &   0.00000 &   Sec   &    I    &  OGLE III  \\
 OGLE-LMC-ECL-13620 &   53299.74348 &   0.00096 &   Prim  &    I    &  OGLE III  \\
 OGLE-LMC-ECL-13620 &   53651.03161 &   0.00286 &   Sec   &    I    &  OGLE III  \\
 OGLE-LMC-ECL-13620 &   54000.04313 &   0.00530 &   Prim  &    I    &  OGLE III  \\
 OGLE-LMC-ECL-13620 &   54001.18487 &   0.00439 &   Sec   &    I    &  OGLE III  \\
 OGLE-LMC-ECL-13620 &   54439.84054 &   0.00235 &   Prim  &    I    &  OGLE III  \\
 OGLE-LMC-ECL-13620 &   54440.99195 &   0.00181 &   Sec   &    I    &  OGLE III  \\
 OGLE-LMC-ECL-13620 &   54839.07964 &   0.00064 &   Prim  &    I    &  OGLE III  \\
 OGLE-LMC-ECL-13620 &   54840.24687 &   0.00434 &   Sec   &    I    &  OGLE III  \\
 OGLE-LMC-ECL-13620 &   57028.63689 &   0.00100 &   Sec   &    R    &  DK154     \\
 OGLE-LMC-ECL-13620 &   57031.73783 &   0.00077 &   Prim  &    R    &  DK154     \\   \hline
 OGLE-LMC-ECL-13666 &   48800.45764 &   0.00368 &   Prim  &    B+R  &  MACHO     \\
 OGLE-LMC-ECL-13666 &   48802.12942 &   0.00774 &   Sec   &    B+R  &  MACHO     \\
 OGLE-LMC-ECL-13666 &   49799.96551 &   0.00587 &   Prim  &    B+R  &  MACHO     \\
 OGLE-LMC-ECL-13666 &   49801.60936 &   0.00445 &   Sec   &    B+R  &  MACHO     \\
 OGLE-LMC-ECL-13666 &   50199.76844 &   0.00428 &   Prim  &    B+R  &  MACHO     \\
 OGLE-LMC-ECL-13666 &   50201.39370 &   0.00526 &   Sec   &    B+R  &  MACHO     \\
 OGLE-LMC-ECL-13666 &   51199.27717 &   0.00314 &   Prim  &    B+R  &  MACHO     \\
 OGLE-LMC-ECL-13666 &   51200.86972 &   0.00441 &   Sec   &    B+R  &  MACHO     \\
 OGLE-LMC-ECL-13666 &   52225.89852 &   0.00053 &   Prim  &    I    &  OGLE III  \\
 OGLE-LMC-ECL-13666 &   52227.43622 &   0.00429 &   Sec   &    I    &  OGLE III  \\
 OGLE-LMC-ECL-13666 &   52598.59894 &   0.00198 &   Prim  &    I    &  OGLE III  \\
 OGLE-LMC-ECL-13666 &   52600.12301 &   0.00248 &   Sec   &    I    &  OGLE III  \\
 OGLE-LMC-ECL-13666 &   52923.86521 &   0.00127 &   Prim  &    I    &  OGLE III  \\
 OGLE-LMC-ECL-13666 &   52925.37570 &   0.00231 &   Sec   &    I    &  OGLE III  \\
 OGLE-LMC-ECL-13666 &   53299.94817 &   0.00921 &   Prim  &    I    &  OGLE III  \\
 OGLE-LMC-ECL-13666 &   53301.45429 &   0.00248 &   Sec   &    I    &  OGLE III  \\
 OGLE-LMC-ECL-13666 &   53648.93441 &   0.00176 &   Prim  &    I    &  OGLE III  \\
 OGLE-LMC-ECL-13666 &   53650.42244 &   0.00254 &   Sec   &    I    &  OGLE III  \\
 OGLE-LMC-ECL-13666 &   54001.30010 &   0.00202 &   Prim  &    I    &  OGLE III  \\
 OGLE-LMC-ECL-13666 &   54002.78112 &   0.00102 &   Sec   &    I    &  OGLE III   \\
 OGLE-LMC-ECL-13666 &   54441.76027 &   0.00204 &   Prim  &    I    &  OGLE III   \\
 OGLE-LMC-ECL-13666 &   54443.22696 &   0.00230 &   Sec   &    I    &  OGLE III   \\
 OGLE-LMC-ECL-13666 &   54841.56466 &   0.00163 &   Prim  &    I    &  OGLE III   \\
 OGLE-LMC-ECL-13666 &   54843.01890 &   0.00274 &   Sec   &    I    &  OGLE III   \\
 OGLE-LMC-ECL-13666 &   56264.58338 &   0.00038 &   Prim  &    R    &  DK154      \\
 OGLE-LMC-ECL-13666 &   56972.70509 &   0.00036 &   Prim  &    R    &  DK154      \\
 OGLE-LMC-ECL-13666 &   57031.73211 &   0.00053 &   Sec   &    R    &  DK154      \\   \hline
 OGLE-LMC-ECL-14771 &   52280.64360 &   0.00255 &   Prim  &    I    &  OGLE III   \\
 OGLE-LMC-ECL-14771 &   52281.65623 &   0.00277 &   Sec   &    I    &  OGLE III   \\
 OGLE-LMC-ECL-14771 &   52622.96379 &   0.00107 &   Prim  &    I    &  OGLE III   \\
 OGLE-LMC-ECL-14771 &   52624.00532 &   0.00084 &   Sec   &    I    &  OGLE III   \\
  \noalign{\smallskip}\hline
\end{tabular}
\end{minipage}
\end{table}

\begin{table}[b]
 \centering
  \begin{minipage}{95mm}
 \fontsize{1.8mm}{2.4mm}\selectfont
 \caption{List of the minima timings used for the analysis - cont.}
\begin{tabular}{ccclcl}
\hline\hline\noalign{\smallskip}
 Star       &    JD Hel.- &  Error & Type   &  Filter  & Source /     \\
            &   2400000   &  [day] &        &          & Observatory  \\
\noalign{\smallskip}\hline
\noalign{\smallskip}
 OGLE-LMC-ECL-14771 &   52984.91102 &   0.00124 &   Prim  &    I    &  OGLE III   \\
 OGLE-LMC-ECL-14771 &   52985.98211 &   0.00222 &   Sec   &    I    &  OGLE III   \\
 OGLE-LMC-ECL-14771 &   53287.99846 &   0.00545 &   Prim  &    I    &  OGLE III   \\
 OGLE-LMC-ECL-14771 &   53289.08411 &   0.00166 &   Sec   &    I    &  OGLE III   \\
 OGLE-LMC-ECL-14771 &   53687.00004 &   0.00496 &   Prim  &    I    &  OGLE III   \\
 OGLE-LMC-ECL-14771 &   53688.12111 &   0.00131 &   Sec   &    I    &  OGLE III   \\
 OGLE-LMC-ECL-14771 &   53998.78656 &   0.00190 &   Prim  &    I    &  OGLE III   \\
 OGLE-LMC-ECL-14771 &   53999.95047 &   0.00505 &   Sec   &    I    &  OGLE III   \\
 OGLE-LMC-ECL-14771 &   54476.29771 &   0.00287 &   Prim  &    I    &  OGLE III   \\
 OGLE-LMC-ECL-14771 &   54477.47995 &   0.00272 &   Sec   &    I    &  OGLE III   \\
 OGLE-LMC-ECL-14771 &   54820.80242 &   0.00488 &   Prim  &    I    &  OGLE III   \\
 OGLE-LMC-ECL-14771 &   54822.00975 &   0.01181 &   Sec   &    I    &  OGLE III   \\
 OGLE-LMC-ECL-14771 &   49057.99767 &   0.00844 &   Prim  &    B+R  &  MACHO       \\
 OGLE-LMC-ECL-14771 &   49058.79013 &   0.00421 &   Sec   &    B+R  &  MACHO       \\
 OGLE-LMC-ECL-14771 &   49350.17760 &   0.00353 &   Prim  &    B+R  &  MACHO       \\
 OGLE-LMC-ECL-14771 &   49350.98966 &   0.00400 &   Sec   &    B+R  &  MACHO       \\
 OGLE-LMC-ECL-14771 &   49648.89471 &   0.00193 &   Prim  &    B+R  &  MACHO       \\
 OGLE-LMC-ECL-14771 &   49649.72701 &   0.00617 &   Sec   &    B+R  &  MACHO       \\
 OGLE-LMC-ECL-14771 &   49945.43060 &   0.00449 &   Prim  &    B+R  &  MACHO       \\
 OGLE-LMC-ECL-14771 &   49946.28295 &   0.00343 &   Sec   &    B+R  &  MACHO       \\
 OGLE-LMC-ECL-14771 &   50250.68726 &   0.00517 &   Prim  &    B+R  &  MACHO       \\
 OGLE-LMC-ECL-14771 &   50251.55429 &   0.00289 &   Sec   &    B+R  &  MACHO       \\
 OGLE-LMC-ECL-14771 &   50597.36981 &   0.00402 &   Prim  &    B+R  &  MACHO       \\
 OGLE-LMC-ECL-14771 &   50598.25695 &   0.00446 &   Sec   &    B+R  &  MACHO       \\
 OGLE-LMC-ECL-14771 &   51175.18171 &   0.00263 &   Prim  &    B+R  &  MACHO       \\
 OGLE-LMC-ECL-14771 &   51176.11505 &   0.00576 &   Sec   &    B+R  &  MACHO       \\     \hline
 OGLE-LMC-ECL-15895 &   52286.83982 &   0.00395 &   Prim  &    I    &  OGLE III    \\
 OGLE-LMC-ECL-15895 &   52288.35649 &   0.01024 &   Sec   &    I    &  OGLE III    \\
 OGLE-LMC-ECL-15895 &   52622.87535 &   0.00658 &   Prim  &    I    &  OGLE III    \\
 OGLE-LMC-ECL-15895 &   52624.38414 &   0.00427 &   Sec   &    I    &  OGLE III    \\
 OGLE-LMC-ECL-15895 &   52985.97548 &   0.00674 &   Prim  &    I    &  OGLE III    \\
 OGLE-LMC-ECL-15895 &   52987.48739 &   0.00599 &   Sec   &    I    &  OGLE III    \\
 OGLE-LMC-ECL-15895 &   53289.49664 &   0.00734 &   Prim  &    I    &  OGLE III    \\
 OGLE-LMC-ECL-15895 &   53291.02978 &   0.00679 &   Sec   &    I    &  OGLE III    \\
 OGLE-LMC-ECL-15895 &   53698.68683 &   0.00312 &   Prim  &    I    &  OGLE III    \\
 OGLE-LMC-ECL-15895 &   53700.20491 &   0.01472 &   Sec   &    I    &  OGLE III    \\
 OGLE-LMC-ECL-15895 &   53999.46999 &   0.00098 &   Prim  &    I    &  OGLE III    \\
 OGLE-LMC-ECL-15895 &   54000.96672 &   0.02676 &   Sec   &    I    &  OGLE III    \\
 OGLE-LMC-ECL-15895 &   54476.43833 &   0.00346 &   Prim  &    I    &  OGLE III    \\
 OGLE-LMC-ECL-15895 &   54477.93117 &   0.00358 &   Sec   &    I    &  OGLE III    \\
 OGLE-LMC-ECL-15895 &   54820.58754 &   0.00527 &   Prim  &    I    &  OGLE III    \\
 OGLE-LMC-ECL-15895 &   54822.08051 &   0.01040 &   Sec   &    I    &  OGLE III    \\
 OGLE-LMC-ECL-15895 &   50520.00247 &   0.00258 &   Prim  &    I    &  OGLE II     \\
 OGLE-LMC-ECL-15895 &   50521.50625 &   0.00583 &   Sec   &    I    &  OGLE II     \\
 OGLE-LMC-ECL-15895 &   50812.67954 &   0.00443 &   Prim  &    I    &  OGLE II     \\
 OGLE-LMC-ECL-15895 &   50814.18421 &   0.01134 &   Sec   &    I    &  OGLE II     \\
 OGLE-LMC-ECL-15895 &   51189.34664 &   0.00738 &   Prim  &    I    &  OGLE II     \\
 OGLE-LMC-ECL-15895 &   51190.84344 &   0.00350 &   Sec   &    I    &  OGLE II     \\
 OGLE-LMC-ECL-15895 &   51641.89855 &   0.00324 &   Prim  &    I    &  OGLE II     \\
 OGLE-LMC-ECL-15895 &   51643.40644 &   0.00566 &   Sec   &    I    &  OGLE II     \\
 OGLE-LMC-ECL-15895 &   49013.32079 &   0.01350 &   Prim  &    B+R  &  MACHO       \\
 OGLE-LMC-ECL-15895 &   49014.81532 &   0.00909 &   Sec   &    B+R  &  MACHO       \\
 OGLE-LMC-ECL-15895 &   49349.34333 &   0.01711 &   Prim  &    B+R  &  MACHO       \\
 OGLE-LMC-ECL-15895 &   49350.85120 &   0.01012 &   Sec   &    B+R  &  MACHO       \\
 OGLE-LMC-ECL-15895 &   49650.14314 &   0.00794 &   Prim  &    B+R  &  MACHO       \\
 OGLE-LMC-ECL-15895 &   49651.63787 &   0.02090 &   Sec   &    B+R  &  MACHO       \\
 OGLE-LMC-ECL-15895 &   49948.22828 &   0.01090 &   Prim  &    B+R  &  MACHO       \\
 OGLE-LMC-ECL-15895 &   49949.72329 &   0.01092 &   Sec   &    B+R  &  MACHO       \\
 OGLE-LMC-ECL-15895 &   50249.01942 &   0.00658 &   Prim  &    B+R  &  MACHO       \\
 OGLE-LMC-ECL-15895 &   50250.51965 &   0.00989 &   Sec   &    B+R  &  MACHO        \\
 OGLE-LMC-ECL-15895 &   50601.30330 &   0.01736 &   Prim  &    B+R  &  MACHO        \\
 OGLE-LMC-ECL-15895 &   50602.80940 &   0.01151 &   Sec   &    B+R  &  MACHO        \\
 OGLE-LMC-ECL-15895 &   51173.08349 &   0.00920 &   Prim  &    B+R  &  MACHO        \\
 OGLE-LMC-ECL-15895 &   51174.59363 &   0.02641 &   Sec   &    B+R  &  MACHO        \\
 OGLE-LMC-ECL-15895 &   56347.72008 &   0.00123 &   Sec   &    R    &  DK154        \\
 OGLE-LMC-ECL-15895 &   56705.39995 &   0.00449 &   Sec   &    R    &  DK154        \\
 OGLE-LMC-ECL-15895 &   56939.75894 &   0.00638 &   Prim  &    R    &  DK154        \\
 OGLE-LMC-ECL-15895 &   57053.55327 &   0.00195 &   Prim  &    R    &  DK154        \\
 OGLE-LMC-ECL-15895 &   57057.67081 &   0.00079 &   Sec   &    R    &  DK154        \\     \hline
 OGLE-LMC-ECL-18102 &   48725.10261 &   0.00390 &   Prim  &    B+R  &  MACHO        \\
 OGLE-LMC-ECL-18102 &   48726.53712 &   0.01530 &   Sec   &    B+R  &  MACHO        \\
 OGLE-LMC-ECL-18102 &   49574.49099 &   0.00429 &   Prim  &    B+R  &  MACHO         \\
 OGLE-LMC-ECL-18102 &   49575.92120 &   0.00278 &   Sec   &    B+R  &  MACHO         \\
 OGLE-LMC-ECL-18102 &   49900.66311 &   0.00189 &   Prim  &    B+R  &  MACHO         \\
 OGLE-LMC-ECL-18102 &   49902.08830 &   0.00155 &   Sec   &    B+R  &  MACHO         \\
 OGLE-LMC-ECL-18102 &   50299.31289 &   0.00219 &   Prim  &    B+R  &  MACHO         \\
 OGLE-LMC-ECL-18102 &   50300.73554 &   0.00620 &   Sec   &    B+R  &  MACHO         \\
 OGLE-LMC-ECL-18102 &   51250.63467 &   0.00279 &   Prim  &    B+R  &  MACHO         \\
 OGLE-LMC-ECL-18102 &   51252.04220 &   0.00398 &   Sec   &    B+R  &  MACHO         \\
 OGLE-LMC-ECL-18102 &   52224.63003 &   0.00000 &   Prim  &    I    &  OGLE III      \\
  \noalign{\smallskip}\hline
\end{tabular}
\end{minipage}
\end{table}

\begin{table}[b]
 \centering
  \begin{minipage}{95mm}
 \fontsize{1.8mm}{2.4mm}\selectfont
 \caption{List of the minima timings used for the analysis - cont.}
\begin{tabular}{ccclcl}
\hline\hline\noalign{\smallskip}
 Star       &    JD Hel.- &  Error & Type   &  Filter  & Source /     \\
            &   2400000   &  [day] &        &          & Observatory  \\
\noalign{\smallskip}\hline
\noalign{\smallskip}
 OGLE-LMC-ECL-18102 &   52225.98281 &   0.00112 &   Sec   &    I   &   OGLE III  \\
 OGLE-LMC-ECL-18102 &   52600.63429 &   0.00138 &   Prim  &    I   &   OGLE III  \\
 OGLE-LMC-ECL-18102 &   52601.96959 &   0.00100 &   Sec   &    I   &   OGLE III  \\
 OGLE-LMC-ECL-18102 &   52924.54314 &   0.00155 &   Prim  &    I   &   OGLE III  \\
 OGLE-LMC-ECL-18102 &   52925.85973 &   0.00128 &   Sec   &    I   &   OGLE III  \\
 OGLE-LMC-ECL-18102 &   53300.54319 &   0.00194 &   Prim  &    I   &   OGLE III  \\
 OGLE-LMC-ECL-18102 &   53301.84519 &   0.00500 &   Sec   &    I   &   OGLE III  \\
 OGLE-LMC-ECL-18102 &   53649.36942 &   0.00131 &   Prim  &    I   &   OGLE III  \\
 OGLE-LMC-ECL-18102 &   53650.64646 &   0.00238 &   Sec   &    I   &   OGLE III  \\
 OGLE-LMC-ECL-18102 &   54000.45929 &   0.00204 &   Prim  &    I   &   OGLE III  \\
 OGLE-LMC-ECL-18102 &   54001.71805 &   0.00186 &   Sec   &    I   &   OGLE III  \\
 OGLE-LMC-ECL-18102 &   54439.88847 &   0.00481 &   Prim  &    I   &   OGLE III  \\
 OGLE-LMC-ECL-18102 &   54441.12600 &   0.00324 &   Sec   &    I   &   OGLE III  \\
 OGLE-LMC-ECL-18102 &   54840.81291 &   0.00313 &   Prim  &    I   &   OGLE III  \\
 OGLE-LMC-ECL-18102 &   54842.01847 &   0.00301 &   Sec   &    I   &   OGLE III  \\
 OGLE-LMC-ECL-18102 &   56942.82917 &   0.00035 &   Prim  &    R   &   DK154     \\
 OGLE-LMC-ECL-18102 &   56974.53643 &   0.00291 &   Prim  &    R   &   DK154     \\
 OGLE-LMC-ECL-18102 &   56975.58552 &   0.00150 &   Sec   &    R   &   DK154     \\      \hline
 OGLE-LMC-ECL-19759 &   48983.32857 &   0.00344 &   Prim  &    B+R &   MACHO     \\
 OGLE-LMC-ECL-19759 &   48984.92795 &   0.01066 &   Sec   &    B+R &   MACHO     \\
 OGLE-LMC-ECL-19759 &   49171.67416 &   0.00677 &   Prim  &    B+R &   MACHO     \\
 OGLE-LMC-ECL-19759 &   49173.27577 &   0.01916 &   Sec   &    B+R &   MACHO     \\
 OGLE-LMC-ECL-19759 &   49428.76274 &   0.00465 &   Prim  &    B+R &   MACHO     \\
 OGLE-LMC-ECL-19759 &   49430.37499 &   0.03134 &   Sec   &    B+R &   MACHO     \\
 OGLE-LMC-ECL-19759 &   49676.88127 &   0.00274 &   Prim  &    B+R &   MACHO     \\
 OGLE-LMC-ECL-19759 &   49678.50924 &   0.00738 &   Sec   &    B+R &   MACHO     \\
 OGLE-LMC-ECL-19759 &   49924.99918 &   0.00623 &   Prim  &    B+R &   MACHO     \\
 OGLE-LMC-ECL-19759 &   49926.63793 &   0.00859 &   Sec   &    B+R &   MACHO     \\
 OGLE-LMC-ECL-19759 &   50176.12105 &   0.00286 &   Prim  &    B+R &   MACHO     \\
 OGLE-LMC-ECL-19759 &   50177.76446 &   0.00552 &   Sec   &    B+R &   MACHO     \\
 OGLE-LMC-ECL-19759 &   50495.98508 &   0.00632 &   Prim  &    B+R &   MACHO     \\
 OGLE-LMC-ECL-19759 &   50497.63632 &   0.00584 &   Sec   &    B+R &   MACHO     \\
 OGLE-LMC-ECL-19759 &   50848.74883 &   0.00451 &   Prim  &    B+R &   MACHO     \\
 OGLE-LMC-ECL-19759 &   50850.42920 &   0.01383 &   Sec   &    B+R &   MACHO     \\
 OGLE-LMC-ECL-19759 &   51273.25886 &   0.01056 &   Prim  &    B+R &   MACHO     \\
 OGLE-LMC-ECL-19759 &   51274.93496 &   0.00907 &   Sec   &    B+R &   MACHO     \\
 OGLE-LMC-ECL-19759 &   50845.76286 &   0.00109 &   Prim  &    I   &   OGLE II   \\
 OGLE-LMC-ECL-19759 &   50847.41129 &   0.00250 &   Sec   &    I   &   OGLE II   \\
 OGLE-LMC-ECL-19759 &   51204.49224 &   0.00202 &   Prim  &    I   &   OGLE II   \\
 OGLE-LMC-ECL-19759 &   50847.41129 &   0.00250 &   Sec   &    I   &   OGLE II   \\
 OGLE-LMC-ECL-19759 &   51551.26748 &   0.00036 &   Prim  &    I   &   OGLE II   \\
 OGLE-LMC-ECL-19759 &   51552.95453 &   0.00475 &   Sec   &    I   &   OGLE II   \\
 OGLE-LMC-ECL-19759 &   51551.26748 &   0.00036 &   Prim  &    I   &   OGLE II   \\
 OGLE-LMC-ECL-19759 &   51827.97770 &   0.00321 &   Sec   &    I   &   OGLE II   \\
 OGLE-LMC-ECL-19759 &   52289.67922 &   0.00211 &   Prim  &    I   &   OGLE III  \\
 OGLE-LMC-ECL-19759 &   52291.36545 &   0.00179 &   Sec   &    I   &   OGLE III  \\
 OGLE-LMC-ECL-19759 &   52630.48662 &   0.00161 &   Prim  &    I   &   OGLE III  \\
 OGLE-LMC-ECL-19759 &   52632.15632 &   0.00179 &   Sec   &    I   &   OGLE III  \\
 OGLE-LMC-ECL-19759 &   52630.48662 &   0.00161 &   Prim  &    I   &   OGLE III  \\
 OGLE-LMC-ECL-19759 &   52987.91647 &   0.00049 &   Sec   &    I   &   OGLE III  \\
 OGLE-LMC-ECL-19759 &   53288.16629 &   0.00049 &   Prim  &    I   &   OGLE III  \\
 OGLE-LMC-ECL-19759 &   53289.85318 &   0.00520 &   Sec   &    I   &   OGLE III  \\
 OGLE-LMC-ECL-19759 &   53697.73384 &   0.00234 &   Prim  &    I   &   OGLE III  \\
 OGLE-LMC-ECL-19759 &   53699.43085 &   0.00597 &   Sec   &    I   &   OGLE III  \\
 OGLE-LMC-ECL-19759 &   53999.67901 &   0.00368 &   Prim  &    I   &   OGLE III  \\
 OGLE-LMC-ECL-19759 &   54001.35087 &   0.00251 &   Sec   &    I   &   OGLE III  \\
 OGLE-LMC-ECL-19759 &   54480.99372 &   0.00120 &   Prim  &    I   &   OGLE III  \\
 OGLE-LMC-ECL-19759 &   54482.65377 &   0.00117 &   Sec   &    I   &   OGLE III  \\
 OGLE-LMC-ECL-19759 &   54818.80260 &   0.00238 &   Prim  &    I   &   OGLE III  \\
 OGLE-LMC-ECL-19759 &   54820.46188 &   0.00774 &   Sec   &    I   &   OGLE III  \\
 OGLE-LMC-ECL-19759 &   56975.85980 &   0.00057 &   Sec   &    R   &   DK154     \\
 OGLE-LMC-ECL-19759 &   56999.78156 &   0.00134 &   Sec   &    R   &   DK154     \\
 OGLE-LMC-ECL-19759 &   57029.67427 &   0.00071 &   Sec   &    R   &   DK154     \\
 OGLE-LMC-ECL-19759 &   57050.60130 &   0.00174 &   Sec   &    R   &   DK154     \\
 OGLE-LMC-ECL-19759 &   57053.58734 &   0.00165 &   Sec   &    R   &   DK154     \\
 OGLE-LMC-ECL-19759 &   57056.57743 &   0.00077 &   Sec   &    R   &   DK154     \\          \hline
 OGLE-LMC-ECL-20112 &   48983.05231 &   0.00453 &   Prim  &    B+R &   MACHO     \\
 OGLE-LMC-ECL-20112 &   48984.64237 &   0.00392 &   Sec   &    B+R &   MACHO     \\
 OGLE-LMC-ECL-20112 &   49171.68324 &   0.00317 &   Prim  &    B+R &   MACHO     \\
 OGLE-LMC-ECL-20112 &   49173.26482 &   0.01156 &   Sec   &    B+R &   MACHO     \\
 OGLE-LMC-ECL-20112 &   49425.24699 &   0.00303 &   Prim  &    B+R &   MACHO     \\
 OGLE-LMC-ECL-20112 &   49426.82362 &   0.00284 &   Sec   &    B+R &   MACHO     \\
 OGLE-LMC-ECL-20112 &   49675.72555 &   0.00278 &   Prim  &    B+R &   MACHO     \\
 OGLE-LMC-ECL-20112 &   49677.29340 &   0.00233 &   Sec   &    B+R &   MACHO     \\
 OGLE-LMC-ECL-20112 &   49926.20303 &   0.00295 &   Prim  &    B+R &   MACHO     \\
 OGLE-LMC-ECL-20112 &   49927.75999 &   0.00305 &   Sec   &    B+R &   MACHO     \\
 OGLE-LMC-ECL-20112 &   50176.68045 &   0.00317 &   Prim  &    B+R &   MACHO     \\
 OGLE-LMC-ECL-20112 &   50178.22672 &   0.00158 &   Sec   &    B+R &   MACHO     \\
 OGLE-LMC-ECL-20112 &   50495.18467 &   0.00473 &   Prim  &    B+R &   MACHO     \\
 OGLE-LMC-ECL-20112 &   50496.72310 &   0.00328 &   Sec   &    B+R &   MACHO     \\
  \noalign{\smallskip}\hline
\end{tabular}
\end{minipage}
\end{table}

\begin{table}[b]
 \centering
  \begin{minipage}{95mm}
 \fontsize{1.8mm}{2.4mm}\selectfont
 \caption{List of the minima timings used for the analysis - cont.}
\begin{tabular}{ccclcl}
\hline\hline\noalign{\smallskip}
 Star       &    JD Hel.- &  Error & Type   &  Filter  & Source /     \\
            &   2400000   &  [day] &        &          & Observatory  \\
\noalign{\smallskip}\hline
\noalign{\smallskip}
 OGLE-LMC-ECL-20112 &   50847.70940 &   0.00502 &   Prim  &    B+R  &  MACHO       \\
 OGLE-LMC-ECL-20112 &   50849.23561 &   0.00285 &   Sec   &    B+R  &  MACHO       \\
 OGLE-LMC-ECL-20112 &   51274.45833 &   0.00477 &   Prim  &    B+R  &  MACHO       \\
 OGLE-LMC-ECL-20112 &   51275.96449 &   0.00397 &   Sec   &    B+R  &  MACHO       \\
 OGLE-LMC-ECL-20112 &   52291.83887 &   0.00136 &   Prim  &    I    &  OGLE III    \\
 OGLE-LMC-ECL-20112 &   52293.31348 &   0.00089 &   Sec   &    I    &  OGLE III    \\
 OGLE-LMC-ECL-20112 &   52631.98245 &   0.00285 &   Prim  &    I    &  OGLE III    \\
 OGLE-LMC-ECL-20112 &   52633.44871 &   0.00174 &   Sec   &    I    &  OGLE III    \\
 OGLE-LMC-ECL-20112 &   52987.59048 &   0.00142 &   Prim  &    I    &  OGLE III    \\
 OGLE-LMC-ECL-20112 &   52989.04614 &   0.00083 &   Sec   &    I    &  OGLE III    \\
 OGLE-LMC-ECL-20112 &   53278.25823 &   0.00198 &   Prim  &    I    &  OGLE III    \\
 OGLE-LMC-ECL-20112 &   53279.71282 &   0.00051 &   Sec   &    I    &  OGLE III    \\
 OGLE-LMC-ECL-20112 &   53698.79902 &   0.00081 &   Prim  &    I    &  OGLE III    \\
 OGLE-LMC-ECL-20112 &   53700.24835 &   0.00097 &   Sec   &    I    &  OGLE III    \\
 OGLE-LMC-ECL-20112 &   54001.83603 &   0.00136 &   Prim  &    I    &  OGLE III    \\
 OGLE-LMC-ECL-20112 &   54003.28330 &   0.00116 &   Sec   &    I    &  OGLE III    \\
 OGLE-LMC-ECL-20112 &   54481.12794 &   0.00143 &   Prim  &    I    &  OGLE III    \\
 OGLE-LMC-ECL-20112 &   54482.56576 &   0.00084 &   Sec   &    I    &  OGLE III    \\
 OGLE-LMC-ECL-20112 &   54818.17657 &   0.00135 &   Prim  &    I    &  OGLE III    \\
 OGLE-LMC-ECL-20112 &   54819.62277 &   0.00158 &   Sec   &    I    &  OGLE III    \\
 OGLE-LMC-ECL-20112 &   56971.82498 &   0.00186 &   Sec   &    R    &  DK154       \\
 OGLE-LMC-ECL-20112 &   56999.65565 &   0.00120 &   Sec   &    R    &  DK154       \\
 OGLE-LMC-ECL-20112 &   57050.74241 &   0.00610 &   Prim  &    R    &  DK154       \\
 OGLE-LMC-ECL-20112 &   57053.82091 &   0.00016 &   Prim  &    R    &  DK154       \\
 OGLE-LMC-ECL-20112 &   57056.92344 &   0.00158 &   Prim  &    R    &  DK154       \\       \hline
 OGLE-LMC-ECL-20438 &   52292.70968 &   0.00485 &   Prim  &    I    &  OGLE III    \\
 OGLE-LMC-ECL-20438 &   52294.27894 &   0.01191 &   Sec   &    I    &  OGLE III    \\
 OGLE-LMC-ECL-20438 &   52634.04974 &   0.00320 &   Prim  &    I    &  OGLE III    \\
 OGLE-LMC-ECL-20438 &   52635.64122 &   0.00240 &   Sec   &    I    &  OGLE III    \\
 OGLE-LMC-ECL-20438 &   52985.65362 &   0.00407 &   Prim  &    I    &  OGLE III    \\
 OGLE-LMC-ECL-20438 &   52987.21342 &   0.00607 &   Sec   &    I    &  OGLE III    \\
 OGLE-LMC-ECL-20438 &   53279.19563 &   0.00404 &   Prim  &    I    &  OGLE III    \\
 OGLE-LMC-ECL-20438 &   53280.77250 &   0.00884 &   Sec   &    I    &  OGLE III    \\
 OGLE-LMC-ECL-20438 &   53695.62695 &   0.00235 &   Prim  &    I    &  OGLE III    \\
 OGLE-LMC-ECL-20438 &   53697.20161 &   0.00432 &   Sec   &    I    &  OGLE III    \\
 OGLE-LMC-ECL-20438 &   53999.42707 &   0.00329 &   Prim  &    I    &  OGLE III    \\
 OGLE-LMC-ECL-20438 &   54000.95887 &   0.00231 &   Sec   &    I    &  OGLE III    \\
 OGLE-LMC-ECL-20438 &   54480.73398 &   0.00331 &   Prim  &    I    &  OGLE III    \\
 OGLE-LMC-ECL-20438 &   54482.28889 &   0.00476 &   Sec   &    I    &  OGLE III    \\
 OGLE-LMC-ECL-20438 &   54818.66145 &   0.00217 &   Prim  &    I    &  OGLE III    \\
 OGLE-LMC-ECL-20438 &   54820.22326 &   0.00472 &   Sec   &    I    &  OGLE III    \\
 OGLE-LMC-ECL-20438 &   49183.07030 &   0.01084 &   Prim  &    B+R  &  MACHO       \\
 OGLE-LMC-ECL-20438 &   49184.68404 &   0.00815 &   Sec   &    B+R  &  MACHO       \\
 OGLE-LMC-ECL-20438 &   49572.19783 &   0.00422 &   Prim  &    B+R  &  MACHO       \\
 OGLE-LMC-ECL-20438 &   49573.80179 &   0.01102 &   Sec   &    B+R  &  MACHO       \\
 OGLE-LMC-ECL-20438 &   49899.89672 &   0.01767 &   Prim  &    B+R  &  MACHO       \\
 OGLE-LMC-ECL-20438 &   49901.49699 &   0.00923 &   Sec   &    B+R  &  MACHO       \\
 OGLE-LMC-ECL-20438 &   50299.27687 &   0.01249 &   Prim  &    B+R  &  MACHO       \\
 OGLE-LMC-ECL-20438 &   50300.88550 &   0.00990 &   Sec   &    B+R  &  MACHO       \\
 OGLE-LMC-ECL-20438 &   51026.32485 &   0.00779 &   Prim  &    B+R  &  MACHO       \\
 OGLE-LMC-ECL-20438 &   51027.92363 &   0.00699 &   Sec   &    B+R  &  MACHO       \\
 OGLE-LMC-ECL-20438 &   56352.83824 &   0.00112 &   Sec   &    R    &  DK154       \\
 OGLE-LMC-ECL-20438 &   57035.55385 &   0.00621 &   Sec   &    R    &  DK154       \\
 OGLE-LMC-ECL-20438 &   57057.84055 &   0.00282 &   Prim  &    R    &  DK154       \\       \hline
 OGLE-LMC-ECL-20498 &   52292.02945 &   0.00194 &   Prim  &    I    &  OGLE III    \\
 OGLE-LMC-ECL-20498 &   52293.10471 &   0.00282 &   Sec   &    I    &  OGLE III    \\
 OGLE-LMC-ECL-20498 &   52634.04620 &   0.00104 &   Prim  &    I    &  OGLE III    \\
 OGLE-LMC-ECL-20498 &   52635.12586 &   0.00109 &   Sec   &    I    &  OGLE III    \\
 OGLE-LMC-ECL-20498 &   52986.43240 &   0.00067 &   Prim  &    I    &  OGLE III    \\
 OGLE-LMC-ECL-20498 &   52987.51029 &   0.00239 &   Sec   &    I    &  OGLE III    \\
 OGLE-LMC-ECL-20498 &   53278.69882 &   0.00240 &   Prim  &    I    &  OGLE III    \\
 OGLE-LMC-ECL-20498 &   53279.77549 &   0.00098 &   Sec   &    I    &  OGLE III    \\
 OGLE-LMC-ECL-20498 &   53693.27394 &   0.00034 &   Prim  &    I    &  OGLE III    \\
 OGLE-LMC-ECL-20498 &   53694.35155 &   0.00170 &   Sec   &    I    &  OGLE III    \\
 OGLE-LMC-ECL-20498 &   54000.05373 &   0.00044 &   Prim  &    I    &  OGLE III    \\
 OGLE-LMC-ECL-20498 &   54001.14180 &   0.00141 &   Sec   &    I    &  OGLE III    \\
 OGLE-LMC-ECL-20498 &   54480.95241 &   0.00075 &   Prim  &    I    &  OGLE III    \\
 OGLE-LMC-ECL-20498 &   54482.03172 &   0.00166 &   Sec   &    I    &  OGLE III    \\
 OGLE-LMC-ECL-20498 &   54818.83122 &   0.00227 &   Prim  &    I    &  OGLE III    \\
 OGLE-LMC-ECL-20498 &   54819.90409 &   0.00268 &   Sec   &    I    &  OGLE III    \\
 OGLE-LMC-ECL-20498 &   48983.78048 &   0.00446 &   Prim  &    B+R  &  MACHO       \\
 OGLE-LMC-ECL-20498 &   48984.79031 &   0.00810 &   Sec   &    B+R  &  MACHO       \\
 OGLE-LMC-ECL-20498 &   49170.33789 &   0.00215 &   Prim  &    B+R  &  MACHO       \\
 OGLE-LMC-ECL-20498 &   49171.36065 &   0.00315 &   Sec   &    B+R  &  MACHO       \\
 OGLE-LMC-ECL-20498 &   49427.38354 &   0.00802 &   Prim  &    B+R  &  MACHO       \\
 OGLE-LMC-ECL-20498 &   49428.38902 &   0.00686 &   Sec   &    B+R  &  MACHO       \\
 OGLE-LMC-ECL-20498 &   49045.96676 &   0.00443 &   Prim  &    B+R  &  MACHO       \\
 OGLE-LMC-ECL-20498 &   49046.98074 &   0.00923 &   Sec   &    B+R  &  MACHO       \\
 OGLE-LMC-ECL-20498 &   49363.12242 &   0.01029 &   Prim  &    B+R  &  MACHO       \\
 OGLE-LMC-ECL-20498 &   49364.13217 &   0.00674 &   Sec   &    B+R  &  MACHO       \\
  \noalign{\smallskip}\hline
\end{tabular}
\end{minipage}
\end{table}

\begin{table}[b]
 \centering
  \begin{minipage}{95mm}
 \fontsize{1.8mm}{2.4mm}\selectfont
 \caption{List of the minima timings used for the analysis - cont.}
\begin{tabular}{ccclcl}
\hline\hline\noalign{\smallskip}
 Star       &    JD Hel.- &  Error & Type   &  Filter  & Source /     \\
            &   2400000   &  [day] &        &          & Observatory  \\
\noalign{\smallskip}\hline
\noalign{\smallskip}
 OGLE-LMC-ECL-20498 &   49674.03832 &   0.00403 &   Prim   &   B+R  &  MACHO   \\
 OGLE-LMC-ECL-20498 &   49675.06246 &   0.00707 &   Sec    &   B+R  &  MACHO   \\
 OGLE-LMC-ECL-20498 &   49924.85282 &   0.00437 &   Prim   &   B+R  &  MACHO   \\
 OGLE-LMC-ECL-20498 &   49925.87046 &   0.00409 &   Sec    &   B+R  &  MACHO   \\
 OGLE-LMC-ECL-20498 &   50177.72492 &   0.01448 &   Prim   &   B+R  &  MACHO   \\
 OGLE-LMC-ECL-20498 &   50178.77624 &   0.00507 &   Sec    &   B+R  &  MACHO   \\
 OGLE-LMC-ECL-20498 &   50494.88176 &   0.00465 &   Prim   &   B+R  &  MACHO   \\
 OGLE-LMC-ECL-20498 &   50495.92021 &   0.00758 &   Sec    &   B+R  &  MACHO   \\
 OGLE-LMC-ECL-20498 &   50849.32928 &   0.00309 &   Prim   &   B+R  &  MACHO   \\
 OGLE-LMC-ECL-20498 &   50850.37785 &   0.00712 &   Sec    &   B+R  &  MACHO   \\
 OGLE-LMC-ECL-20498 &   51276.33528 &   0.00283 &   Prim   &   B+R  &  MACHO   \\
 OGLE-LMC-ECL-20498 &   51277.38675 &   0.00321 &   Sec    &   B+R  &  MACHO   \\
 OGLE-LMC-ECL-20498 &   56974.60825 &   0.00245 &   Prim   &   R    &  DK154    \\
 OGLE-LMC-ECL-20498 &   57034.71784 &   0.00306 &   Prim   &   R    &  DK154    \\
 OGLE-LMC-ECL-20498 &   57035.73851 &   0.01039 &   Sec    &   R    &  DK154    \\
 OGLE-LMC-ECL-20498 &   57057.52572 &   0.00039 &   Prim   &   R    &  DK154    \\
 OGLE-LMC-ECL-20498 &   57059.59527 &   0.00175 &   Prim   &   R    &  DK154    \\
   \noalign{\smallskip}\hline
\end{tabular}
\end{minipage}
\end{table}

\end{appendix}
\end{document}